\documentclass[twocolumn,superscriptaddress,floatfix,preprintnumbers]{revtex4}
\usepackage{graphics,amssymb,amsmath,epsfig,color}
\usepackage{graphicx}

\begin{document}
\title{Resilient quantum electron microscopy}
\author{Hiroshi Okamoto}
\affiliation{Department of Intelligent Mechatronics, Akita Prefectural University, Yurihonjo 015-0055, Japan}
\date{\today}
\begin{abstract}
We investigate the fundamental limit of biological quantum electron microscopy, which is designed to go beyond the shot noise limit. Inelastic scattering is expected to be the main obstacle in this setting, especially for thick specimens of actual biological interest. Here we describe a measurement procedure that, in principle, significantly neutralizes the effect of inelastic scattering.
\end{abstract}
\maketitle

\section{Introduction}

The raw resolution of biological electron cryomicroscopy (cryoEM)
is manifestly limited by shot noise. This is due to the small number
of imaging electrons intended for avoiding radiation damage to the
frozen specimen \cite{henderson_review}. In single particle analysis
(SPA), for instance, the tolerable number of electrons, i.e. the electron
\emph{fluence}, is at most $\approx5\times10^{3}/\mathrm{nm^{2}}$
\cite{grant grigorieff}. On the other hand, biological objects are
weak phase objects. Hence shot noise tends to bury the signal and
thus limits the attainable resolution.

Quantum metrology, where phase measurement is a standard problem,
is a natural approach to improving cryoEM. Recall that measuring a
small phase $\theta$ with precision $\delta\theta$ takes $N\approx\delta\theta^{-2}$
electrons because of the shot noise limit. There have recently been
proposals of quantum electron microscopy (QEM) schemes for approaching
the Heisenberg limit, where $N\approx\delta\theta^{-1}$. Some quantum
schemes are based on repeated use of single electrons \cite{Bob_phys_today,putnam_yanik,Sebastian_Thomas,kruit_qem,multipass_tem,Agarwal_etal,linear_cavity,Koppell_10kV},
while others use entanglement between electrons and superconducting
qubits \cite{eeem_cpb,eeem_flux,eeem_error}. Nonetheless, many of
these methods accumulate the small phase $\theta$ onto a quantum
object $k$ times, resulting in a phase $k\theta$ after $k$ electron-passing
events through the specimen, which is then measured. Call this process
a single \emph{round} of measurement, and call $k$ the \emph{repetition
number}. This is equivalent to measuring a hypothetical object with
associated phase shift $k\theta$, using hypothetical $N/k$ probe
particles at the shot noise limit. As a result, we obtain an increased
effective number of electrons as $kN\approx\delta\theta^{-2}$, which
approaches the Heisenberg limit at $k=N$. However, the usable value
of $k$ depends on the frequency of inelastic scattering. If inelastic
scattering destroys a round of measurement, then all electron passages
used in that round are wasted.

In this work, we explore the limit of QEM. We ask a question, ``Can
we neutralize the adverse effect of inelastic scattering at least
partially?'', and give an affirmative answer. To explore the physical
limit, as opposed to the engineering limit, we assume the full ability
to manipulate and measure the combined system of imaging electrons
and other quantum objects. Without loss of generality, ``other quantum
objects'' may be thought of as a set of qubits. In short, we consider
an electron microscope connected to a quantum computer, which we may
call a universal QEM \cite{q_interface}.

The \emph{raw} resolution of current cryoEM is about $3-5\,\mathrm{nm}$
\cite{raw resolution cryoET}. All high resolution data to date are
obtained only by averaging over at least tens of thousands of molecules
of the same structure, by using e.g. SPA. However, the biologist would
ultimately want to see molecules in their cellular context, rather
than as ensemble average of purified molecules. We focus on unique,
single specimens in the present work. At present, only very large
proteins ($\sim$MDa) are identifiable in electron cryotomography
(ECT) \cite{baumeister_review}. High energy electrons generally are
desirable in ECT to ensure transmission of electrons especially when
the specimen is tilted. Moreover, the \emph{effective} thickness of
the specimen is $k$ times the actual thickness in QEM. Hence, hereafter
we focus on $300\,\mathrm{keV}$ electrons with the wavelength $\lambda=1.97\,\mathrm{pm}$.

The specimen thickness $t$ is an important parameter in QEM. As quantum
measurement is limited by lossy events, a relevant length to be compared
with $t$ is the inelastic mean free path $\Lambda=200-350\,\mathrm{nm}$
for $300\,\mathrm{keV}$ electrons \cite{vulovic_etal}. Suppose,
for now, that \emph{all} inelastic scattering destroy quantum measurement.
The fraction of quantum measurements that survive to the end is $e^{-kt/\Lambda}$
because of $k$ electron passing events. Hence we replace the number
of rounds $N/k$ with $Ne^{-kt/\Lambda}/k$. We thus modify the above
relation $kN\approx\delta\theta^{-2}$ to $kNe^{-kt/\Lambda}\approx\delta\theta^{-2}$.
The optimal $k$ that maximizes $\delta\theta^{-2}$ is $k_{1}=\Lambda/t$,
where we have a relation $\delta\theta\approx\sqrt{e/k_{1}N}$. Improvement
over the shot noise limit $\delta\theta\approx1/\sqrt{N}$ in terms
of the phase measurement precision is therefore 
\begin{equation}
\sqrt{\frac{k_{1}}{e}}=\sqrt{\frac{\Lambda}{et}}\approx\sqrt{\frac{100\,\mathrm{nm}}{t}}.\label{eq:garden_variety_QEM_improvement}
\end{equation}
This result emphasizes the importance of thinning the specimen. However,
perhaps $t$ cannot be smaller than the size of biological molecules,
e.g. $\approx10\,\mathrm{nm}$. Moreover, in cryoEM of vitreous sections
(CEMOVIS), the specimen thickness is ``rarely less than $50\,\mathrm{nm}$''
\cite{CEMOVIS sample thickness}. Hence, to attain sizable improvement,
we must \emph{neutralize} the effect of inelastic scattering, which
is the central topic of this paper.

We list some conventions. We generally represent the length of a vector,
for example $\mathbf{a}$, using the same symbol, i.e. $a=|\mathbf{a}|$.
Symbols $\hat{\mathbf{i}},\,\hat{\mathbf{j}},\,\hat{\mathbf{k}}$
denote unit vectors parallel to $x,\,y$ and $z$ axes, respectively.
The electron optical axis is $z$. A position in real space is represented
as $\mathbf{r}=x\hat{\mathbf{i}}+y\hat{\mathbf{j}}+z\hat{\mathbf{k}}$.
A wave vector in the reciprocal space is written as $\mathbf{k}=k_{x}\hat{\mathbf{i}}+k_{y}\hat{\mathbf{j}}+k_{z}\hat{\mathbf{k}}$.
Its size is related to the wavelength $\lambda$ as $k=2\pi/\lambda$,
as opposed to the crystallographic definition $k=1/\lambda$. In many
cases, we only need projections of vectors onto the $xy$-plane, which
are represented by the same symbols when there is no danger of confusion.
Eigenstates of the position and momentum operators are $|\mathbf{r}\rangle$
and $|\mathbf{k}\rangle$, respectively. We will often use $2$-dimensional
(2d) Fourier transform (FT) and its inverse: 
\begin{equation}
F\left(\mathbf{k}\right)=\mathcal{F}_{C}\left\{ f\left(\mathbf{r}\right)\right\} =\int f\left(\mathbf{r}\right)e^{-i\mathbf{k}\cdot\mathbf{r}}d^{2}\mathbf{r},\label{eq:cont_FT_S8}
\end{equation}
\begin{equation}
f\left(\mathbf{r}\right)=\mathcal{F}_{C}^{-1}\left\{ F\left(\mathbf{k}\right)\right\} =\int F\left(\mathbf{k}\right)e^{i\mathbf{k}\cdot\mathbf{r}}\frac{d^{2}\mathbf{k}}{\left(2\pi\right)^{2}},
\end{equation}
where the subscript $C$ denotes ``continuous'' of continuous FT.
When there is no risk of confusion, a tensor product $|a\rangle\otimes|b\rangle$
is simply written as $|a\rangle|b\rangle$ or $|ab\rangle$. The rest
mass of the electron is denoted as $m_{e}$. We use conventional relativistic
notations $\beta=\frac{v}{c}$ and $\gamma=\frac{1}{\sqrt{1-\beta^{2}}}$.
Additional conventions are presented at relevant places.

\section{Resolution-dependent specimen damage\label{sec:specimen_damage}}

Specimen damage starts from short-range structural features progressively
towards long-range features. A recent transmission electron microscopy
(TEM) study \cite{Russo} on a purple membrane 2d crystal describes
radiation damage in a way that is particularly amenable to theoretical
analysis. Let the scattering vector be $\mathbf{q}=\mathbf{k}_{f}-\mathbf{k}_{i}$,
where $\mathbf{k}_{i},\,\mathbf{k}_{f}$ are electron wave vectors
before and after scattering. The vector $\mathbf{q}$ is practically
perpendicular to the optical axis $z$ and we will occasionally treat
$\mathbf{q}$ as a $2$d vector in the $q_{x}q_{y}$-plane. The intensity
of the electron wave scattered off the crystal at a diffraction plane
is found to decay as \cite{Russo}

\begin{equation}
I=I_{0}e^{-RFq^{2}/8\pi^{2}},
\end{equation}
where $I$ is the intensity, $I_{0}$ is the initial intensity, $R\approx7\times10^{-4}\mathrm{nm}^{4}$
is a constant, $F$ is electron fluence. The initial intensity $I_{0}$
is proportional to $\varSigma_{0}e^{-R_{g}^{2}q^{2}/3}$, as usually
found in the Guinier plot, where $R_{g}$ is the radius of gyration
of the molecule under study \cite{rosenthal_henderson}. However,
at higher spatial frequencies, scattered waves from atoms interfere
essentially at random, giving constant average intensity with respect
to $\mathbf{q}$ that is the sum of intensities from each atom, with
random phase. Although the particular value of $R$ above pertains
to the purple membrane, we assume that the value generalizes fairly
well to other proteins.

The ``$B$-factor'' $B=RF$ has a natural interpretation that electron
irradiation basically causes random walks of atoms, recalling that
the standard $B$-factor in X-ray crystallography expresses the square
of thermal atomic displacements. Indeed, we show that $\sqrt{B/8\pi^{2}}$
may be regarded as the expected positional deviation of atoms from
the original location. Let the position and the electron scattering
amplitude (having the dimension of length) of $s$-th atom be $\mathbf{r}_{s}$
and $f_{s}$, respectively. Our argument is valid to the extent that
$f_{s}$ can be regarded as a constant within the range of scattering
angle of interest. The scattered electron wavefunction amplitude $\psi\left(\mathbf{q}\right)$
in the far field is proportional to 
\begin{equation}
\psi\left(\mathbf{q}\right)\propto\sum_{s}f_{s}e^{-i\mathbf{q}\cdot\mathbf{r}_{s}}.
\end{equation}
In the present case the scattering vector $\mathbf{q}$ lies almost
exactly in a plane perpendicular to the optical axis. Hence we treat
$\mathbf{q}$, as well as $\mathbf{r}_{s}$, as 2-dimensional. Introducing
a function $\gamma\left(\mathbf{r}\right)$ representing the projected
``density of the scattering amplitude'', we obtain 
\begin{equation}
\psi\left(\mathbf{q}\right)\propto\int d^{2}\mathbf{r}\gamma\left(\mathbf{r}\right)e^{-i\mathbf{q}\cdot\mathbf{r}}.
\end{equation}
Suppose that atoms random walk. We model this process by convoluting
the projected density of scattering amplitude $\gamma\left(\mathbf{r}\right)$
with 
\begin{equation}
g\left(\mathbf{r}\right)=\frac{1}{2\pi d^{2}}e^{-\frac{x^{2}+y^{2}}{2d^{2}}}
\end{equation}
upon irradiation, where $d$ is the standard positional displacement
from the initial positions of atoms. Fourier transforming, we obtain
the scattering amplitude that is $\psi\left(\mathbf{q}\right)$ multiplied
by $\mathcal{F}_{C}\left\{ g\left(\mathbf{r}\right)\right\} =e^{-\frac{d^{2}q^{2}}{2}}$.
Hence the initial intensity $I_{0}=\left|\psi\left(\mathbf{q}\right)\right|^{2}$
at zero radiation damage is multiplied by $e^{-d^{2}q^{2}}$. This
allows us to identify $d^{2}$ with $B/8\pi^{2}$.

Finally, we note a limitation of this approach. We implicitly assumed
that there are \emph{many} atoms random-walking so that $\gamma\left(\mathbf{r}\right)$
may be considered to be convoluted with a gaussian function. After
a long time, all the intensity on the diffraction plane is concentrated
at $\mathbf{q}=0$ in this model. However, all the atoms should remain
at some particular positions rather than having smoothed out by gaussian
averaging. Hence random intensity in the far field should remain.

\section{The measurement procedure\label{sec:The-measurement-procedure}}

\subsection{High-level ideas\label{subsec:High-level-ideas}}

We begin with a high-level description of our measurement procedure.
Since small structural features disappear fast, it is sensible to
\emph{selectively} acquire high spatial frequency (SF) data first.
Selective-SF measurement makes sense also in view of inelastic scattering
because high-SF measurements, associated with high-angle scattering,
tend to be insensitive to small-angle inelastic scattering for reasons
to be described later. Hence our strategy is to repeat measurement
of scattered-wave amplitude at a specific SF, beginning at a large
$q$ region and progressively moving inwards in the far field. We
show that universal QEM, irrespective of the physical system it is
based on, allows for such selective-SF measurement. However, selective-SF
measurement may most easily be understood as an extended version of
entanglement-enhanced electron microscopy (EEEM, see Appendix A) based
on superconducting qubits.

We suppress unwanted signals outside the chosen SF by \emph{not} performing
phase-to-amplitude conversion and by \emph{not} quantum-enhance them.
As discussed in the previous section, unwanted low-resolution signals
tend to be larger than the high-resolution signal we are after in
typical biological specimens. Such unwanted large signal adversely
interfere with high resolution measurement because the relation between
the phase shift to contrast is not exactly linear in phase contrast
microscopy.

The allowable fluence at each SF is rigidly constrained. Let $A$
be the area of electron beam illumination and define $\sigma=\frac{\pi}{q}$
as the resolution of interest. Since $B/8\pi^{2}=RF/8\pi^{2}$ is
essentially the mean squared distance traveled by random-walking atoms
under electron irradiation, structural information on the length scale
$\sigma$ should be obtained before $\sigma^{2}\approx B/8\pi^{2}$,
i.e. electron fluence $F$ reaches $F_{\mathrm{opt}}=\zeta\frac{8\pi^{2}\sigma^{2}}{R}$,
where $\zeta=O\left(1\right)$ is a numerical constant.

We briefly digress to give an argument that gives $\zeta=0.255$ as
a plausibly optimal numerical value. The overall physical picture
is that too small a fluence $F$ gives no statistical confidence,
while too large an $F$ yields data that mostly reflect altered structures
due to radiation damage. Hence an optimal $F_{\mathrm{opt}}$ should
exist. To find a useful $\zeta$ value, we make a pragmatic assumption
that a Fourier component $\theta_{\mathbf{q}}$ of the real-space
map of weak phase shift $\theta\left(\mathbf{r}\right)$ of the specimen
decays to \emph{zero} as $\theta_{\mathbf{q}}\left(F\right)=\theta_{0}e^{-F/F_{0}}$
for some $F_{0}$. Strictly speaking, this assumption cannot be entirely
right (see the last paragraph of Sec. \ref{sec:specimen_damage})
but we hope to obtain a useful estimate nonetheless. As shown in the
quantitative study of radiation damage \cite{Russo}, $\theta_{\mathbf{q}}^{2}$
is proportional to the diffraction intensity $I\left(\mathbf{q}\right)$
and hence 
\begin{equation}
\theta_{\mathbf{q}}\left(F\right)=\theta_{0}e^{-RFq^{2}/16\pi^{2}}.
\end{equation}
It follows that $F_{0}=\frac{16\pi^{2}}{Rq^{2}}=\frac{16\sigma^{2}}{R}\approx2.3\times10^{4}\,\mathrm{nm}^{-2}\cdot\left(\sigma/\mathrm{nm}\right)^{2}$.
This is sufficiently large and unless we are after very high resolution
data, we can ignore the specimen change \emph{during} each round of
quantum measurement, assuming the repetition number of the order of
$k\approx\Lambda/t\approx10$. Let $|s\rangle$ be the unscattered
electron state and $|a\rangle$ be the scattered state with the wave
vector $\mathbf{q}$. Deferring the question of how to perform SF-selective
measurement, in principle we obtain a quantum state $|s\rangle+ik\theta_{\mathbf{q}}\left(F\right)|a\rangle$
after a quantum-enhanced measurement with the repetition number $k$,
providing $k\theta_{\mathbf{q}}\left(F\right)\ll1$ (Also see later
discussions in this subsection). By expressing the state with measurement
basis states $\left|\uparrow\right\rangle =\left[\left(1+i\right)|s\rangle+\left(1-i\right)|a\rangle\right]/2$
and $\left|\downarrow\right\rangle =\left[\left(1-i\right)|s\rangle+\left(1+i\right)|a\rangle\right]/2$,
we obtain the corresponding probabilities $p_{\uparrow}=\frac{1}{2}-k\theta_{\mathbf{q}}\left(F\right)$
and $p_{\downarrow}=\frac{1}{2}+k\theta_{\mathbf{q}}\left(F\right)$.
Let $X$ be a random variable that represents the number of events
``$\uparrow$'' occurring after $N_{g}=AF/k$ quantum enhanced measurements
on a specimen area $A$, each using a group of $k$ electrons. Note
that $p_{\uparrow}$ is a function of $F$ and hence that of $N_{g}$,
because the specimen gradually gets damaged. The expectation value
$\overline{X}$ is given by 
\[
\overline{X}=\int p_{\uparrow}\left(N_{g}\right)dN_{g}\approx\frac{N_{g}}{2}-k\theta_{0}\int_{0}^{N_{g}}e^{-kN'_{g}/AF_{0}}dN'_{g}
\]
\begin{equation}
=\frac{N_{g}}{2}-AF_{0}\theta_{0}\left(1-e^{-kN_{g}/AF_{0}}\right),
\end{equation}
while the variance is approximately a constant with respect to $\theta_{0}$,
i.e. 
\begin{equation}
\mathrm{Var}\left(X\right)=\int_{0}^{N_{g}}p_{\uparrow}\left(N_{g}'\right)p_{\downarrow}\left(N_{g}'\right)dN_{g}'\approx\frac{N_{g}}{4}.
\end{equation}
The estimator for $\theta_{0}$ is 
\begin{equation}
\hat{\theta}_{0}\left(X\right)=\frac{\nu/k}{1-e^{-\nu}}\left(\frac{1}{2}-\frac{X}{N_{g}}\right),
\end{equation}
where $\nu=kN_{g}/AF_{0}=F/F_{0}$. We obtain 
\begin{equation}
\mathrm{Var}\left(\hat{\theta}_{0}\left(X\right)\right)\approx\frac{1}{4kAF_{0}}\frac{\nu}{\left(1-e^{-\nu}\right)^{2}},
\end{equation}
which is minimized at $\nu_{\mathrm{opt}}\approx1.26$, where $e^{\nu}=2\nu+1$
is satisfied. Hence we obtain $\zeta=2F_{\mathrm{opt}}/\left(\pi^{2}F_{0}\right)=2\nu_{\mathrm{opt}}/\pi^{2}\approx0.255$.

Having found an appropriate value of $\zeta$, we proceed to consider
our highly constrained way to spend the fluence budget to each SF
bands. The electron fluence that one can expend in a ring-shaped resolution
band $\left[q,q+\Delta q\right]$ on the $q_{x}q_{y}$ plane is 
\[
\Delta F=F_{\mathrm{opt}}\left(q\right)-F_{\mathrm{opt}}\left(q+\Delta q\right)
\]
\begin{equation}
\approx-\frac{dF_{\mathrm{opt}}\left(q\right)}{dq}\Delta q=\zeta\frac{16\pi^{4}}{Rq^{3}}\Delta q.
\end{equation}
There are $\frac{2\pi q}{\Delta q}$ square-shaped regions with the
side length $\Delta q$ in the band $\left[q,q+\Delta q\right]$ in
the $q$-space. Thus, the fluence budget for the measurement at each
square-shaped region is $F_{\mathrm{sq}}=\frac{\Delta F}{2\pi q/\Delta q}$.
A natural scale of $\Delta q$ satisfies $A\cdot\Delta q^{2}=\left(2\pi\right)^{2}$,
where $A$ is the imaging area in the real space, since we do not
have structures finer than $\Delta q$ in the reciprocal space. Thus,
we are allowed to spend electron dose 
\begin{equation}
N_{\mathrm{sq}}=F_{\mathrm{sq}}A=\zeta\frac{32\pi\sigma^{4}}{R}=3.7\times10^{4}\left(\sigma/\mathrm{nm}\right)^{4}\label{eq:quartic_dose}
\end{equation}
for measuring scattered wave amplitude at a small area $\Delta q^{2}$
in the $q$-space. This expression has a \emph{quartic} dependence
on $\sigma$, meaning that allowed fluence is much smaller at a higher
SF. Note that Eq. (\ref{eq:quartic_dose}) does not depend on the
area $A$. The reason is that a large area in the real space is associated
with a finer $\Delta q$ and hence many points need to be scanned
in the reciprocal space.

In the rest of this section, we briefly sketch the method of acquiring
data at a specific SF $\mathbf{q}$. To focus on the essence of the
idea, consider 1-dimensional specimen and we write $\mathbf{q}=q\hat{\mathbf{i}}=\frac{\pi}{\sigma}\hat{\mathbf{i}}$.
The specimen is a weak phase object and an incident wave $e^{ik_{z}z}$
is scattered into a state 
\begin{equation}
e^{i\left[k_{z}z+\theta\left(x\right)\right]}\approx e^{ik_{z}z}\left[1+i\theta\left(x\right)\right],
\end{equation}
where $\theta\left(x\right)\ll1$ is the phase shift map that we want
to determine. It is natural to assume that the process is insensitive
to the tilt of the incident wave, and hence for a small $k$ we have
\begin{equation}
e^{i\left[k_{z}z+kx+\theta\left(x\right)\right]}\approx e^{i\left(k_{z}z+kx\right)}\left[1+i\theta\left(x\right)\right].
\end{equation}
Henceforth we omit the common factor $e^{ik_{z}z}$. Let the incident
electron state be superposition of plane waves with the $x$ component
of wavevectors separated by $2q$: 
\begin{equation}
\psi_{s}\left(x\right)=\sum_{n\in\mathbb{Z}}e^{2niqx}.
\end{equation}
We will pretend that $n$ runs over all integers for mathematical
convenience. Obviously this is an idealization because the aperture
angle is finite and small in real electron optics. Since we focus
on the SF $q$, we study scattering of the incident wave into a state
\begin{equation}
\psi_{a}\left(x\right)=\sum_{n\in\mathbb{Z}}e^{\left(2n+1\right)iqx},
\end{equation}
which has wavevectors in the midpoints between those in the incident
state. We made the incident waves lattice-like for symmetry reasons,
as will be clear shortly. Note that $\psi_{s}\left(x\right)$ and
$\psi_{a}\left(x\right)$ are both real.

Consider a specimen with $\theta\left(x\right)=\theta_{0}\cos\left(qx\right)$
for now to focus on the SF $q$. A plane wave precisely along the
optical axis scatters into 
\begin{equation}
1+i\theta\left(x\right)\approx1+\frac{i\theta_{0}}{2}\left(e^{iqx}+e^{-iqx}\right)
\end{equation}
Superposing this, and utilized the assumption that the scattering
process is not sensitive to small tilt angles, we see that scattering
makes the following transformation: 
\begin{equation}
\psi_{s}\left(x\right)\Rightarrow\psi_{s}\left(x\right)+i\theta_{0}\psi_{a}\left(x\right),\;\psi_{a}\left(x\right)\Rightarrow\psi_{a}\left(x\right)+i\theta_{0}\psi_{s}\left(x\right).\label{eq:scattering_by_grating}
\end{equation}
An alternative way, which provides a complementary view, to derive
Eq.(\ref{eq:scattering_by_grating}) is the following. First, note
that $\psi_{0}\left(x\right)=\psi_{s}\left(x\right)+\psi_{a}\left(x\right)$
and $\psi_{1}\left(x\right)=\psi_{s}\left(x\right)-\psi_{a}\left(x\right)$
are proportional to $\sum_{n\in\mathbb{Z}}\delta\left(x-2n\sigma\right)$
and $\sum_{n\in\mathbb{Z}}\delta\left(x-2n\sigma-\sigma\right)$,
respectively. (To see this, one may either use a physical argument
or the mathematical identity 
\begin{equation}
\sum_{n\in\mathbb{Z}}e^{2\pi inx}=\sum_{n\in\mathbb{Z}}\delta\left(x-n\right).
\end{equation}
) Hence $\psi_{0}\left(x\right)$ and $\psi_{1}\left(x\right)$ should
respectively receive phase shift $\theta_{0}$ and $-\theta_{0}$
because $\theta\left(x\right)=\theta_{0}\cos\left(qx\right)$. It
follows that $\psi_{0}\left(x\right)\Rightarrow\psi_{0}\left(x\right)+i\theta_{0}\psi_{0}\left(x\right)$
and $\psi_{1}\left(x\right)\Rightarrow\psi_{1}\left(x\right)-i\theta_{0}\psi_{1}\left(x\right)$,
which is consistent with Eq.(\ref{eq:scattering_by_grating}). Starting
with the initial state $\psi_{s}\left(x\right)$, we then repeat the
transformation Eq.(\ref{eq:scattering_by_grating}) for $k$ times.
The final state should be $\psi_{s}\left(x\right)+ik\theta_{0}\psi_{a}\left(x\right)$
if $k\theta_{0}\ll1$.

Next, consider general specimens. Unlike the specimen with the structure
$\theta\left(x\right)=\theta_{0}\cos\left(qx\right)$, they scatter
an incoming plane wave into all directions. Since we want to perform
a selective SF measurement, we wish to confine the quantum state within
the Hilbert subspace $\mathcal{H}$ spanned by $\psi_{s}\left(x\right)$
and $\psi_{a}\left(x\right)$. These two states are proportional to
mutually interleaving rows of dots in the reciprocal $k$-space, namely
$\sum_{n}\delta\left(k-2nq\right)$ for $\psi_{s}\left(x\right)$
and $\sum_{n}\delta\left(k-2nq-q\right)$ for $\psi_{a}\left(x\right)$.
Scattering of the primary wave $\psi_{s}\left(x\right)$ into $\psi_{a}\left(x\right)$
is caused not only by the SF component $\pm q$, but also by those
at $\pm3q,\pm5q,\cdots$. As we learned in Sec. \ref{sec:specimen_damage},
higher-SF components are expected to be generally much smaller and
hence we ignore them.

We begin with a SF-selective procedure with a poor performance, which
is instructive nonetheless. We divide the reciprocal $k$-space into
\emph{cells}, that are intervals $\left[\left(n-\frac{1}{2}\right)q,\left(n+\frac{1}{2}\right)q\right)$,
where $n\in\mathbb{Z}$. In other words, we reorganize $k$ into two
variables $n$ and $-\frac{q}{2}\leq\hat{k}<\frac{q}{2}$, such that
$k=nq+\hat{k}$ and hence $\hat{k}$ indicates the position within
a cell. To remain in $\mathcal{H}$, we measure $\hat{k}$ but leave
$n$ unmeasured. (This is conceptually not much different from measuring
the $x$ coordinate of a particle while leaving its $y,z$ coordinates.
Hence this should in principle be possible.) The measurement outcome
would mostly be $\hat{k}=0$ because of the presence of the intense
primary electron beam. In this case, a superposed state of $\psi_{s}\left(x\right)$
and $\psi_{a}\left(x\right)$ remains intact because both of these
have the same value $\hat{k}=0$, although the parity of $n$ is different
between these. On the other hand, if the measurement result is $\hat{k}\neq0$,
i.e. if elastic scattering takes place, the state such as $\psi_{s}\left(x\right)+ik\theta_{0}\psi_{a}\left(x\right)$
is destroyed. To see this, for example consider the similar amplitudes
in the far field at $k=2nq+\hat{k}$ and $k=\left(2n+1\right)q+\hat{k}$,
where $\hat{k}\approx\frac{q}{2}$. Hence the measurement fails. However,
elastic scattering events take place sufficiently often and we cannot
tolerate such a failure.

To remain in the space $\mathcal{H}$ after elastic scattering, we
\emph{obfuscate} the fact that elastic scattering ever happened. This
is done by burying the scattered waves under the intense primary wave,
by recombining these waves. Details are described in the next subsection,
but what follows are some basic ideas. Since eventually we want to
determine $\theta_{0}$ in the state $\psi_{s}\left(x\right)+ik\theta_{0}\psi_{a}\left(x\right)$,
we do not want to recombine the scattered waves with $\psi_{s}\left(x\right)$
in such a way that $\psi_{s}\left(x\right)$ acquires the imaginary
part. Recombination without such acquisition turns out to be possible
because $\theta\left(x\right)$ is real, which in turn imply certain
symmetry of the wave function in the reciprocal space. We note that
this recombination process must involve a \emph{measurement}, since
the quantum state has unknown components outside the Hilbert subspace
$\mathcal{H}$ but we need to project the state back to $\mathcal{H}$.
As a bonus, such a measurement avoids coherently accumulating unwanted
quantum amplitudes that do not belong the SF of interest. It turns
out that a certain symmetry between $\psi_{s}\left(x\right)$ and
$\psi_{a}\left(x\right)$ is desirable in our operations since we
employ an extra quantum entity, namely a qubit. Hence, the recombination
operation of the waves is performed separately in each cell $\left[n-\frac{q}{2},n+\frac{q}{2}\right)$.
Eventually, we obtain a state of the form 
\begin{equation}
\left(1+A\right)\psi_{s}\left(x\right)+\left(ik\theta_{0}+B\right)\psi_{a}\left(x\right),\label{eq:qubit_state_non_ideal}
\end{equation}
where $A,B$ are real, unknown and small amplitudes coming from the
undesirable scattered waves. We now argue that $A$ and $B$ does
not present a significant problem. Consider the Bloch sphere, wherein
$\psi_{0}\left(x\right)$ and $\psi_{1}\left(x\right)$ are north
and south poles, respectively. To determine the imaginary component
$ik\theta_{0}$, we want to measure the state with the measurement
basis states $\psi_{\pm}\left(x\right)\propto\psi_{0}\left(x\right)\pm i\psi_{1}\left(x\right)$.
Note that $\psi_{\pm}\left(x\right)$, $\psi_{s}\left(x\right)$,
$\psi_{a}\left(x\right)$ and $\psi_{s}\left(x\right)+ik\theta\psi_{a}\left(x\right)$
are all on the equator of the sphere. Hence a measurement with respect
to the basis $\psi_{\pm}\left(x\right)$ yields information about
the \emph{longitude} of the state on the Bloch sphere. Since small
real values of $A,B$ mean a small shift in the \emph{latitude} on
the sphere, this will not significantly affect the measurement using
the basis states $\psi_{\pm}\left(x\right)$.

Finally, we emphasize that the primary benefit of the SF-selective
method discussed above lies in the handling of inelastic scattering.
Since inelastic scattering tends to be associated with small scattering
angles, separation of $\psi_{s}$ and $\psi_{a}$ in the far field
helps us protect the quantum state that we need. See Sec. \ref{sec:Neutralization-of-inelastic}
for further discussions.

\subsection{Low-level procedure\label{subsec:Low-level-procedure}}

We now describe details of the SF-selective measurement sketched above.
Our selective-SF measurement procedure may be understood more smoothly
if the reader knows how the ``conventional'' entanglement-enhanced
electron microscopy (EEEM) works \cite{eeem_cpb,eeem_flux}. See Appendix
A for an introduction to EEEM.

We start with definitions. As in the previous subsection, we consider
an incident electron state that form a lattice of focused beams on
the specimen. However, this time the lattice is a 2d square lattice
with the lattice constant $\sigma$. More specifically, let the number
of focused beams be $M^{2}$, where $M=2^{j}$, where $j\geq2$ is
an integer. We define $k_{\mathrm{max}}=\frac{2\pi}{\sigma}$ and
$k_{\mathrm{min}}=k_{\mathrm{max}}/M$. For later convenience, we
define sets of ordered pairs of integers:
\begin{equation}
\mathcal{M}=\left\{ \left(n,m\right)\in\mathbb{Z}^{2}|-\frac{M}{2}\leq n<\frac{M}{2},-\frac{M}{2}\leq m<\frac{M}{2}\right\} ,
\end{equation}
\begin{equation}
\mathcal{M}_{s}=\left\{ \left(n,m\right)\in\mathcal{M}|n<0\right\} ,
\end{equation}
\begin{equation}
\mathcal{M}_{a}=\left\{ \left(n,m\right)\in\mathcal{M}|0\leq n\right\} ,
\end{equation}
\begin{equation}
\mathcal{M}_{c}=\left\{ \left(n,m\right)\in\mathcal{M}|-\frac{M}{4}\leq n<\frac{M}{4}\right\} ,
\end{equation}
\begin{equation}
\mathcal{M}_{e}=\left\{ \left(n,m\right)\in\mathcal{M}|n\textrm{ is even}\right\} ,
\end{equation}
and
\begin{equation}
\mathcal{M}_{o}=\left\{ \left(n,m\right)\in\mathcal{M}|n\textrm{ is odd}\right\} .
\end{equation}
Subscripts $s,a,c$ are intended to mean ``symmetric'', ``antisymmetric''
and ``center'', respectively. We also define singleton sets $\mathcal{C}_{s}=\left\{ \left(-\frac{M}{4},0\right)\right\} $,
$\mathcal{C}_{a}=\left\{ \left(\frac{M}{4},0\right)\right\} $ and
$\mathcal{C}_{c}=\left\{ \left(0,0\right)\right\} $. The sets $\mathcal{C}_{s}$,
$\mathcal{C}_{a}$ and $\mathcal{C}_{c}$, in a sense, contain the
``central'' element of $\mathcal{M}_{s}$, $\mathcal{M}_{a}$ and
$\mathcal{M}_{c}$, respectively. Now, each focused electron beam
is at $n\sigma\hat{\mathbf{i}}+m\sigma\hat{\mathbf{j}}$, where $\left(n,m\right)\in\mathcal{M}$.
The electron state at the point $n\sigma\hat{\mathbf{i}}+m\sigma\hat{\mathbf{j}}$
is written as $|n,m\rangle$. Since a quantum state can in principle
be transferred, but not copied, between the electron microscope and
a connected quantum computer \cite{q_interface}, the electron state
$|n,m\rangle$ may equivalently be viewed, as we will do occasionally,
as a state of two ``registers of a quantum computer'' $n,\,m$,
each comprising $\log_{2}M$ qubits. They express integers $n,\,m$
in \emph{modified} two's complement notation (MTCN), wherein the sign
bit is reversed: $0$ means negative and $1$ means positive. Let
Q2 be the most significant qubit (MSQ) of the register $n$. The remaining
part of the register $n$ is referred to as register $\tilde{n}$.
The integer $\tilde{n}$ is represented again in MTCN and $-\frac{M}{4}\leq\tilde{n}<\frac{M}{4}$
is satisfied. Hence we have 
\begin{equation}
\tilde{n}=\begin{cases}
n+\frac{M}{4} & \textrm{if}\:-\frac{M}{2}\leq n<0\\
n-\frac{M}{4} & \textrm{if}\:0\leq n<\frac{M}{2}
\end{cases}.\label{eq:tilde_definition}
\end{equation}
In addition to the registers $n,\,m$, we introduce an extra qubit
Q1, which plays a central role.

We denote a state of a qubit with a bar. Let the basis states of a
qubit be $|\overline{0}\rangle,\,|\overline{1}\rangle$. Define $|\overline{s}\rangle=\frac{|\overline{0}\rangle+|\overline{1}\rangle}{\sqrt{2}}$,
$|\overline{a}\rangle=\frac{|\overline{0}\rangle-|\overline{1}\rangle}{\sqrt{2}}$,
$|\overline{\uparrow}\rangle=\frac{|\overline{0}\rangle+i|\overline{1}\rangle}{\sqrt{2}}$
and $|\overline{\downarrow}\rangle=\frac{|\overline{0}\rangle-i|\overline{1}\rangle}{\sqrt{2}}$.
The tensor product of an electron state $|p\rangle$ and the Q1 state
$|\overline{q}\rangle$ will be denoted as $|p\rangle|\overline{q}\rangle$,
or simply $|p\overline{q}\rangle$. Negative of complex conjugate
will be abbreviated as NCC. Let $a_{n,m}$ be a set of $M^{2}$ values,
where $\left(n,m\right)\in\mathcal{M}$. We write 2d discrete FT (DFT)
\begin{equation}
A_{n,m}=\frac{1}{M}\sum_{\left(r,s\right)\in\mathcal{M}}a_{r,s}e^{2\pi i\frac{nr+ms}{M}}\label{eq:DFT}
\end{equation}
as $A_{n,m}=\mathcal{F}\left\{ a_{n,m}\right\} $, where $\left(n,m\right)\in\mathcal{M}$.
The inverse of the DFT is written as $a_{n,m}=\mathcal{F}^{-1}\left\{ A_{n,m}\right\} $.
The DFT is often applied to a set of quantum amplitudes as quantum
FT (QFT \cite{Simon_qFFT}), wherein a state 
\begin{equation}
\sum_{\left(n,m\right)\in\mathcal{M}}a_{n,m}|n,m\rangle\label{eq:q_state_before_FT}
\end{equation}
is converted to 
\begin{equation}
\sum_{\left(n,m\right)\in\mathcal{M}}A_{n,m}|n,m\rangle,\label{eq:Fourier_transformed_q_state}
\end{equation}
where $A_{n,m}=\mathcal{F}\left\{ a_{n,m}\right\} $. Here we further
define \emph{split inverse FT}, which applies inverse-FT \emph{individually}
to two half planes $\mathcal{M}_{s}$ and $\mathcal{M}_{a}$, wherein
the central points are $\left(-\frac{M}{4},0\right)$ and $\left(\frac{M}{4},0\right)$,
respectively. More specifically, $a_{n,m}=\mathcal{F}_{\textrm{split}}^{-1}\left\{ A_{n,m}\right\} $
is a shorthand for 
\begin{equation}
a_{n,m}=\frac{\sqrt{2}}{M}\sum_{\left(r,s\right)\in\mathcal{M}_{s}}A_{r,s}e^{-2\pi i\frac{2\left(r+\frac{M}{4}\right)\left(n+\frac{M}{4}\right)+sm}{M}},
\end{equation}
for $\left(n,m\right)\in\mathcal{M}_{s}$, and 
\begin{equation}
a_{n,m}=\frac{\sqrt{2}}{M}\sum_{\left(r,s\right)\in\mathcal{M}_{a}}A_{r,s}e^{-2\pi i\frac{2\left(r-\frac{M}{4}\right)\left(n-\frac{M}{4}\right)+sm}{M}},
\end{equation}
for $\left(n,m\right)\in\mathcal{M}_{a}$. Alternatively, in accordance
with Eq. (\ref{eq:tilde_definition}) 
\begin{equation}
a_{\tilde{n},m}=\frac{\sqrt{2}}{M}\sum_{\left(r,s\right)\in\mathcal{M}_{s}}A_{r,s}e^{-2\pi i\frac{2\left(r+\frac{M}{4}\right)\tilde{n}+sm}{M}}\label{eq:split_iFT1}
\end{equation}
if $n<0$, and 
\begin{equation}
a_{\tilde{n},m}=\frac{\sqrt{2}}{M}\sum_{\left(r,s\right)\in\mathcal{M}_{a}}A_{r,s}e^{-2\pi i\frac{2\left(r-\frac{M}{4}\right)\tilde{n}+sm}{M}}\label{eq:split_iFT2}
\end{equation}
if $n\ge0$, both for $\left(\tilde{n},m\right)\in\mathcal{M}_{c}$.

Three remarks about the split inverse FT are in order. First, the
prefactor is $\frac{\sqrt{2}}{M}=\frac{1}{\sqrt{\frac{M^{2}}{2}}}$
because the number of elements involved in each FT is $\frac{M^{2}}{2}$.
Second, the exponential factor may be rewritten as, e.g. when $\left(n,m\right)\in\mathcal{M}_{s}$,
\begin{equation}
e^{-2\pi i\frac{2\left(r+\frac{M}{4}\right)\left(n+\frac{M}{4}\right)+sm}{M}}=e^{-2\pi i\frac{\left(r+\frac{M}{4}\right)\left(n+\frac{M}{4}\right)}{M/2}}e^{-2\pi i\frac{sm}{M}}.
\end{equation}
Third, from the QFT perspective, the above operation is equivalent
to performing 2d inverse QFT on the state $|\tilde{n},m\rangle$,
where we set aside the MSQ of the register $n$, i.e. Q2. Hence, we
can define the split inverse QFT as a transformation of the state
$\sum_{\left(\tilde{n},m\right)\in\mathcal{M}_{c}}A_{\tilde{n},m}|\tilde{n},m\rangle$
into $\sum_{\left(\tilde{n},m\right)\in\mathcal{M}_{c}}a_{\tilde{n},m}|\tilde{n},m\rangle$,
where 
\begin{equation}
a_{\tilde{n},m}=\frac{\sqrt{2}}{M}\sum_{\left(\tilde{r},s\right)\in\mathcal{M}_{c}}A_{\tilde{r},s}e^{-2\pi i\frac{2\tilde{r}\tilde{n}+sm}{M}}.
\end{equation}

We consider a thin specimen that is characterized by a 2d phase shift
map $\theta\left(x,y\right)$. The area of the measurement on the
specimen is square-shaped, with the side length $L=M\sigma$. Define
$\theta_{n,m}=\theta\left(n\sigma,m\sigma\right)$. We set 
\begin{equation}
\sum_{\left(n,m\right)\in\mathcal{M}}\theta_{n,m}=0\label{eq:mean_phase_is_zero}
\end{equation}
without loss of generality. We aim to measure 
\begin{equation}
\overline{\theta}=\frac{1}{M^{2}}\sum_{\left(n,m\right)\in\mathcal{M}}\left(-1\right)^{n}\theta_{n,m},\label{eq:theta_bar}
\end{equation}
which obviously contains SF of $q=\frac{\pi}{\sigma}$ as the main
component. That is, half the difference between; (A) the average value
of $\theta_{n,m}$ with even $n$, and (B) the odd $n$ counterpart.
The electron beam array being a \emph{square} lattice is not particularly
important. In addition, the reader may justifiably worry that a focused
beam would quickly destroy the specimen at the focal point. A solution
to this problem is discussed in Appendix B.

We are now ready to discuss the SF-selective measurement procedure.
The electron state $|n,m\rangle$ after transmission through the specimen
is 
\begin{equation}
e^{i\theta_{n,m}}|n,m\rangle\approx\left(1+i\theta_{n,m}\right)|n,m\rangle
\end{equation}
under weak phase approximation, where the small quantity $\theta_{n,m}$
is the phase map of a specimen. Define $\Theta_{n,m}=\mathcal{F}\left\{ \theta_{n,m}\right\} $.
Note that 
\begin{equation}
\Theta_{0,0}=0\label{eq:average_phase_set_to_zero}
\end{equation}
because of Eq. (\ref{eq:mean_phase_is_zero}). Unlike the treatment
in the previous subsection, in the following EEEM-like setting, we
will deal with two symmetrically placed incident states

\begin{equation}
|s\rangle=\frac{1}{M}\sum_{\left(n,m\right)\in\mathcal{M}}e^{i\frac{\pi}{2}n}|n,m\rangle\label{eq:qstate_s}
\end{equation}
and 
\begin{equation}
|a\rangle=\frac{1}{M}\sum_{\left(n,m\right)\in\mathcal{M}}e^{-i\frac{\pi}{2}n}|n,m\rangle,\label{eq:qstate_a}
\end{equation}
which are scattered into each other. (A change of the convention $|n,m\rangle\Rightarrow e^{-i\frac{\pi}{2}n}|n,m\rangle$
would make it better correspond to the ``sketch'' of the method
in Sec. \ref{subsec:High-level-ideas}. However, here we make $|s\rangle$
and $|a\rangle$ more symmetric.) These states interfere to make fringes
at either even $n$ spots or odd ones, as in:

\begin{equation}
|0\rangle=\frac{|s\rangle+|a\rangle}{\sqrt{2}}=\frac{\sqrt{2}}{M}\sum_{\left(n,m\right)\in\mathcal{M}_{e}}\left(-1\right)^{\frac{n}{2}}|n,m\rangle,
\end{equation}
\begin{equation}
|1\rangle=\frac{|s\rangle-|a\rangle}{\sqrt{2}}=\frac{\sqrt{2}i}{M}\sum_{\left(n,m\right)\in\mathcal{M}_{o}}\left(-1\right)^{\frac{n-1}{2}}|n,m\rangle.
\end{equation}
Our intention is to measure the difference of the average phase shifts,
of the transmitted electron beams, between locations belonging to
$|0\rangle$ and $|1\rangle$ on the specimen. DFT transforms states
in Eqs. (\ref{eq:qstate_s}) and (\ref{eq:qstate_a}) into $|-\frac{M}{4},0\rangle$
and $|\frac{M}{4},0\rangle$, respectively.

\begin{widetext}

\begin{figure}[t]
\includegraphics[scale=0.35]{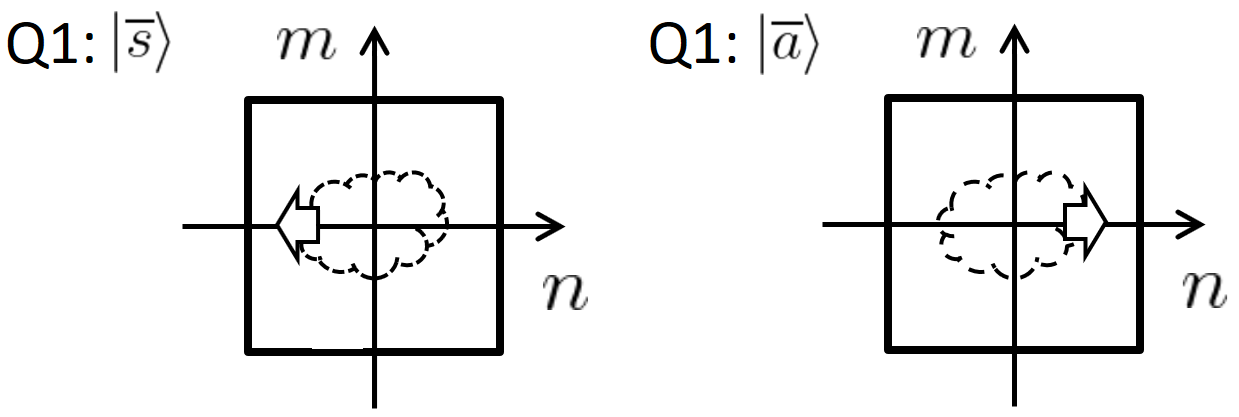}(a)

\includegraphics[scale=0.35]{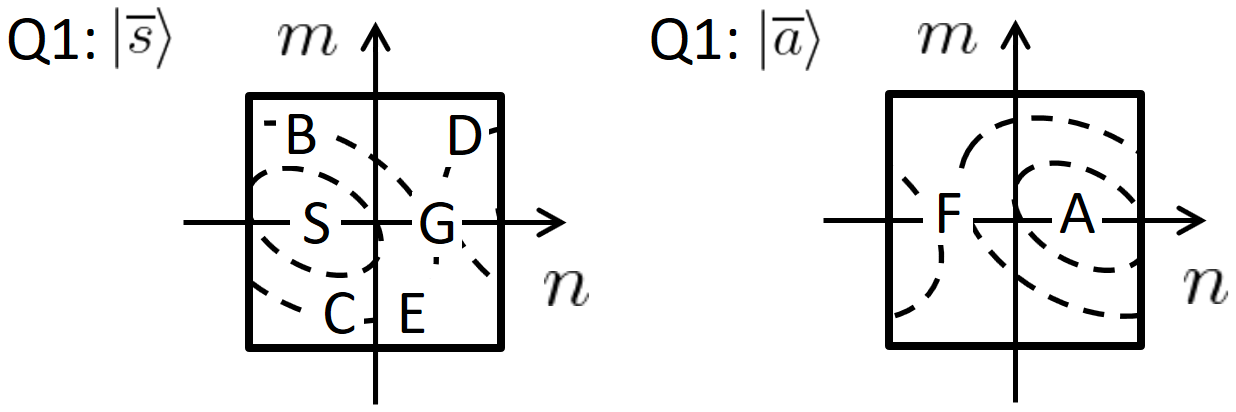}(b)

\includegraphics[scale=0.35]{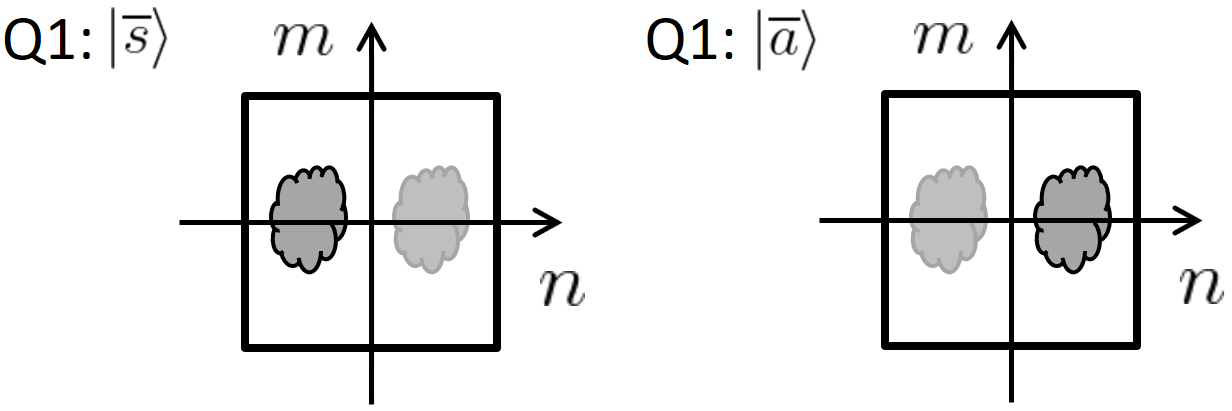}(c)\caption{Illustrations of quantum states at different stages of spatial-frequency
(SF) selective measurement. The electron state, which may have been
transferred to a quantum computer, in the branch of the entire wavefunction,
wherein Q1 state is $|\overline{s}\rangle$, is shown in the left
figure. In a similar way, one with $|\overline{a}\rangle$ is on the
right. See the main text for more information. (a) The state right
after transmission of an electron through the specimen. The shape
of the ``molecule'', which is imprinted on the phase of electron
wavefunction, is shown in dashed curves because it is not ``visible'',
in a loose analogy with transmission electron microscopy (TEM) imaging.
The electron wavefront is such that the wave goes left if Q1 state
is $|\overline{s}\rangle$, and goes right if Q1 state is $|\overline{a}\rangle$,
as indicated by the arrows. (It reflects our convention of discrete
Fourier transform.) (b) We apply QFT to the quantum states $|n,m\rangle$,
representing the transmitted array of $M\times M$ beams, to obtain
essentially a far-field state. The left-moving wave in figure (a)
goes to point S, while the right-moving one goes to A. Elastically
scattered waves, whose intensity is indicated by dashed curves, surround
the transmitted waves at S and A. Unlike actual diffraction in real
electron optics, periodic boundary condition (PBC) is applicable in
quantum Fourier transform (QFT). Thus, the wave that goes beyond the
left boundary, for example, comes out from the right side. The scattered
amplitudes at B and C are the negative of complex conjugate (NCC)
to each other. The same applies to D and E due to PBC. The amplitudes
of interest at F and G are NCC to itself and hence is pure imaginary.
These amplitudes at F and G are ``added'' to, although the wavefunction
branch is different, the transmitted waves respectively at S and A
with a $\frac{\pi}{2}$ phase difference. (c) The state after split
inverse Fourier transformation (FT), where inverse FT is performed
in $n<0$ and $n\ge0$ regions separately. If the outcome of Q2 measurement
in Step $7$ is $c=0$, the $n<0$ part of the map is left available,
while for $c=1$, the $n\ge0$ part remains. In each region, the shape
of the ``molecule'' is recovered to an extent. In contrast to conventional
in-focus phase contrast microscopy, we convert \emph{un}wanted phase
into amplitude in Step $6$ before the split inverse FT, because we
want to accumulate the signal \emph{phase} onto Q1. (In Step $12$,
we measure Q1 with respect to basis states $\left\{ |\overline{\uparrow}\rangle,|\overline{\downarrow}\rangle\right\} $,
thus finally ``converting the phase into amplitude'', figuratively
speaking, when the final measurement is done.) Since the unwanted
information is converted to amplitude and is ``visible'', again
in a loose analogy with TEM, the ``molecules'' are drawn with solid
curves, as opposed to dashed curves that are found in figure (a).
The images of the ``molecule'' is shown dimly when it is high-pass
filtered, where the signal is indeed expected to be weak.}
\end{figure}

\end{widetext}

Figure 1 is designed to visualize aspects of the SF-selective measurement
steps that we describe below. The combined system of the electron
and Q1 has $2M^{2}$ complex amplitudes to specify its state (setting
aside the minor issue of normalization and the overall phase), because
the electron has $M^{2}$ basis vectors $|n,m\rangle$ and Q1 has
$2$ basis vectors $|\overline{s}\rangle,\,|\overline{a}\rangle$.
To visualize the state of the entire system, we can show two maps
Q1:$|\overline{s}\rangle$ and Q1:$|\overline{a}\rangle$, each corresponding
to $|\overline{s}\rangle$ and $|\overline{a}\rangle$ of the Q1 states,
with $n$ and $m$ axes, to show a set of complex coefficients. For
example, the point $\left(n,m\right)$ of the map for Q1:$|\overline{s}\rangle$
shows a complex coefficient, i.e. the quantum amplitude, for the state
$|n,m\rangle|\overline{s}\rangle$. (To visualize complex numbers,
one might use brightness and color to show the amplitude and phase,
respectively, for example. However, we do not do such things in this
paper, because schematic representations of general ideas suffice
for us.)

We first state steps in SF-selective measurement without elucidating
results of performing these steps.

Step $1$: Initialize the state of Q1 to be $|\overline{s}\rangle$.

Step $2$: Initialize the state of a new electron to $\frac{1}{\sqrt{2}}\left(|s\rangle+|a\rangle\right)$.

Step $3$: Apply a CNOT gate that flips Q1 state as $|\overline{s}\rangle\Leftrightarrow|\overline{a}\rangle$
if and only if the electron is in state $|a\rangle$.

Step $4$: Let the electron go through the specimen. Capture the electron
quantum state in registers $n$ and $m$ of a quantum computer. (See
Fig. 1 (a).)

Step $5$: Apply QFT that converts the amplitude of $|n,m\rangle$
from $a_{n,m}$ to $A_{n,m}=\mathcal{F}\left\{ a_{n,m}\right\} $.

Step $6$: Multiply $i$ to two states $|\pm\frac{M}{4},0\rangle$
(i.e. points S, G, F, A in Fig. 1 (b)).

Step $7$: Measure Q2 with respect to basis $\left\{ |\overline{0}\rangle,\,|\overline{1}\rangle\right\} $.
Let the outcome be $c$. (Here we determine if the state is in the
$n<0$, or $n\geq0$, region of the map shown in Fig. 1 (b).)

Step $8$: Apply split inverse QFT. (See Fig. 1 (c). Also note that
Q2 is not involved in this operation.)

Step $9$: Measure the state of the register $\tilde{n}$ (i.e. the
register $n$ without Q2) and $m$ with respect to the basis $\left\{ |\tilde{n},m\rangle|\left(\tilde{n},m\right)\in\mathcal{M}_{c}\right\} $.
(Although the measurement outcome $\hat{n},\,\hat{m}$ contains low-resolution
information, we do not discuss utilization of it in the present work.)

Step $10$: Apply the single-qubit operation $|\overline{s}\rangle\Leftrightarrow|\overline{a}\rangle$
to Q1 if and only if $c=1$.

Step $11$: Go back to step $2$ to repeat the process for $k$ times.

Step $12$: Measure Q1 with the basis states $\left\{ |\overline{\uparrow}\rangle,|\overline{\downarrow}\rangle\right\} $.
One obtains a single bit of data for one round of measurement.

In the rest of this section, we track the state of the microscope
system during the above procedure. Anticipating later arguments, we
write the Q1 state after Step $1$ in a generalized form 
\begin{equation}
|\overline{s}\rangle+i\alpha|\overline{a}\rangle,\label{eq:generalized_initial_state}
\end{equation}
instead of simply writing it as $|\overline{s}\rangle$. Then, as
specified in Step $2$ a new electron is prepared in the state 
\begin{equation}
\frac{1}{\sqrt{2}}\left\{ |s\rangle+|a\rangle\right\} .
\end{equation}
Step $3$ is an entangling operation, which results in 
\begin{equation}
\frac{1}{\sqrt{2}}\left\{ \left(|s\overline{s}\rangle+|a\overline{a}\rangle\right)+i\alpha\left(|s\overline{a}\rangle+|a\overline{s}\rangle\right)\right\} .
\end{equation}
In Step $4$, the electron goes through the specimen. The exit wave
from the specimen generated from the incident wave $|s\rangle$ is
\begin{equation}
|\psi_{s}\rangle=\frac{1}{M}\sum_{\left(n,m\right)\in\mathcal{M}}\left(1+i\theta_{n,m}\right)e^{i\frac{\pi}{2}n}|n,m\rangle.\label{eq:psi_s_d}
\end{equation}
Likewise, for the incident wave $|a\rangle$ we obtain 
\begin{equation}
|\psi_{a}\rangle=\frac{1}{M}\sum_{\left(n,m\right)\in\mathcal{M}}\left(1+i\theta_{n,m}\right)e^{-i\frac{\pi}{2}n}|n,m\rangle.\label{eq:psi_a_d}
\end{equation}
Thus, the state of the entire system, having captured the electron
state in the quantum computer, is 
\begin{equation}
\frac{1}{\sqrt{2}}\left\{ \left(|\psi_{s}\overline{s}\rangle+|\psi_{a}\overline{a}\rangle\right)+i\alpha\left(|\psi_{s}\overline{a}\rangle+|\psi_{a}\overline{s}\rangle\right)\right\} 
\end{equation}
Figure 1 (a) schematically shows the state at this point. Both Q1:$|\overline{s}\rangle$
and Q1:$|\overline{a}\rangle$ maps show the same ``biological molecule''
imprinted as phase shift. However, the ``wave fronts are tilted''
in the opposite direction between these two maps because of the $e^{\pm i\frac{\pi}{2}n}$
factors in Eqs. (\ref{eq:qstate_s}, \ref{eq:qstate_a}). 

Next, in Step $5$, we perform 2d quantum fast Fourier transform (QFT)
\cite{Simon_qFFT}, which converts the amplitude $c_{n,m}$ of the
state $|n,m\rangle$ to $c'_{n,m}=\mathcal{F}\left\{ c_{n,m}\right\} $.
This amounts to moving to the diffraction plane, although the \emph{periodic
boundary condition} (PBC) applies here, unlike in the case of actual
diffraction plane in real electron optics. Note two properties of
Fourier-transformed phase map: 
\begin{equation}
\Theta_{r+M,s}=\Theta_{r,s},\label{eq:periodicity}
\end{equation}
which is the PBC, and 
\begin{equation}
\Theta_{r,s}=\Theta_{-r,-s}^{*},\label{eq:conjugate property}
\end{equation}
because $\theta_{n,m}$ is real. Note that, due to Eq. (\ref{eq:theta_bar}),
\begin{equation}
\overline{\theta}=\frac{\Theta_{\frac{M}{2},0}}{M}=\frac{\Theta_{-\frac{M}{2},0}}{M},\label{eq:theta-bar}
\end{equation}
which is what we aim to measure. QFT applied to $|\psi_{s}\rangle$
and $|\psi_{a}\rangle$ yields, taking Eq. (\ref{eq:average_phase_set_to_zero})
into account, 
\[
|\psi_{s2}\rangle=|-\frac{M}{4},0\rangle+\frac{i}{M}\sum_{\left(r,s\right)\in\mathcal{M}\backslash\mathcal{C}_{s}}\Theta_{r+\frac{M}{4},s}|r,s\rangle
\]
\begin{equation}
=|-\frac{M}{4},0\rangle+i\overline{\theta}|\frac{M}{4},0\rangle+\frac{i}{M}\sum_{\left(r,s\right)\in\mathcal{M}\backslash\left(\mathcal{C}_{s}\cup\mathcal{C}_{a}\right)}\Theta_{r+\frac{M}{4},s}|r,s\rangle,
\end{equation}
and
\[
|\psi_{a2}\rangle=|\frac{M}{4},0\rangle+\frac{i}{M}\sum_{\left(r,s\right)\in\mathcal{M}\backslash\mathcal{C}_{a}}\Theta_{r-\frac{M}{4},s}|r,s\rangle
\]
\begin{equation}
=|\frac{M}{4},0\rangle+i\overline{\theta}|-\frac{M}{4},0\rangle+\frac{i}{M}\sum_{\left(r,s\right)\in\mathcal{M}\backslash\left(\mathcal{C}_{s}\cup\mathcal{C}_{a}\right)}\Theta_{r-\frac{M}{4},s}|r,s\rangle.
\end{equation}
\begin{widetext}The state of the entire system is thus 
\[
|\Psi_{1}\rangle=\frac{1}{\sqrt{2}}|-\frac{M}{4},0\rangle\left[\left(1-\alpha\overline{\theta}\right)|\overline{s}\rangle+i\left(\alpha+\overline{\theta}\right)|\overline{a}\rangle\right]+\frac{1}{\sqrt{2}}|\frac{M}{4},0\rangle\left[\left(1-\alpha\overline{\theta}\right)|\overline{a}\rangle+i\left(\alpha+\overline{\theta}\right)|\overline{s}\rangle\right]
\]
\begin{equation}
+\frac{i}{\sqrt{2}M}\sum_{\left(r,s\right)\in\mathcal{M}\backslash\left(\mathcal{C}_{s}\cup\mathcal{C}_{a}\right)}|r,s\rangle\left\{ \Theta_{r+\frac{M}{4},s}\left(|\overline{s}\rangle+i\alpha|\overline{a}\rangle\right)+\Theta_{r-\frac{M}{4},s}\left(|\overline{a}\rangle+i\alpha|\overline{s}\rangle\right)\right\} .\label{eq:step5}
\end{equation}
\end{widetext}(To verify this, it may help to define and use $|\overline{p}\rangle=|\overline{s}\rangle+i\alpha|\overline{a}\rangle$
and $|\overline{q}\rangle=|\overline{s}\rangle+i\alpha|\overline{a}\rangle$.)
Figure 1 (b) shows this state. In the first term of Eq. (\ref{eq:step5}),
the large component corresponds to the points S, while the smaller
component corresponds to F. A similar statement can be made for the
second term. The transmitted waves of the two ``tilted incident waves''
$|s\rangle,\,|a\rangle$ mainly make two points S and A in the far
field. Elastically scattered waves are in the third term of Eq. (\ref{eq:step5})
and surround these two points. They are shown in dashed curves in
Fig. 1 (b). Because of Eq. (\ref{eq:conjugate property}), points
B and C are NCC to each other. Moreover, the same can be said for
points D and E because of Eq. (\ref{eq:periodicity}).

To motivate the following steps, we first discuss a method of poor
performance. Suppose that we measure the electron state at this point.
If the measurement outcomes were always $n=\pm\frac{M}{4}$ and $m=0$,
then, neglecting $\alpha\overline{\theta}$, we would be able to accumulate
the phase $\overline{\theta}$ on top of $\alpha$ in Q1. However,
sometimes we measure the electron state at other points $\left(r,s\right)$,
when elastic scattering occurs. This would result in a Q1 state 
\begin{equation}
\Theta_{r+\frac{M}{4},s}\left(|\overline{s}\rangle+i\alpha|\overline{a}\rangle\right)+\Theta_{r-\frac{M}{4},s}\left(|\overline{a}\rangle+i\alpha|\overline{s}\rangle\right),
\end{equation}
which is basically a destroyed state unless one of $\Theta_{r+\frac{M}{4},s},\,\Theta_{r-\frac{M}{4},s}$
is overwhelmingly larger than the other, in the sense that the absolute
value of their ratio (or its inverse) has to be much smaller than
$\alpha$, which is generally small to begin with. To avoid such a
scenario, we perform the following steps to obscure the fact that
elastic scattering happened.

Hence in Step $6$, we apply a ``virtual $\pi/2$ phase plate''
by selectively apply a phase factor $i$ to the states $|\pm\frac{M}{4},0\rangle$
that are points S, G, F and A in Fig. 1 (b). (In our subsequent computation,
we simply remove the factor $i$ from the second line of Eq. (\ref{eq:step5}),
yielding an equivalent result.)

Two definitions are in order before we proceed to Steps $7$ and $8$.
We first introduce 
\begin{equation}
\theta_{\tilde{n},m}^{L}=\frac{1}{M}\sum_{\left(r',s\right)\in\mathcal{M}_{c}\backslash\mathcal{C}_{c}}\Theta_{r',s}e^{-2\pi i\frac{2r'\tilde{n}+sm}{M}},\label{eq:theta-low}
\end{equation}
where we impose the range $\left(\tilde{n},m\right)\in\mathcal{M}_{c}$.
The addition of the condition $\left(r',s\right)\notin\mathcal{C}_{c}=\left\{ \left(0,0\right)\right\} $
is unnecessary here, but it makes later arguments clearer. This quantity
clearly represents a low-pass filtered map of $\theta_{\tilde{n},m}$,
which is compressed in the $x$-direction. Because the number of pixels
along the axis of compression is $M/2$, we have the slightly odd-looking
exponent $-2\pi i\frac{r'\tilde{n}}{\left(M/2\right)}-2\pi i\frac{sm}{M}=-2\pi i\frac{2r'\tilde{n}+sm}{M}$.
We also introduce 
\begin{equation}
\theta_{\tilde{n},m}^{H}=\frac{1}{M}\sum_{\left(r',s\right)\in\mathcal{M}_{c}\backslash\mathcal{C}_{c}}\Theta_{r'+\frac{M}{2},s}e^{-2\pi i\frac{2r'\tilde{n}+sm}{M}},\label{eq:theta-high}
\end{equation}
which also has the range $\left(\tilde{n},m\right)\in\mathcal{M}_{c}$.
Note that we subtracted $\overline{\theta}$ by excluding the $\left(r',s\right)=\left(0,0\right)$
term (See Eq. (\ref{eq:theta-bar})). We list several properties of
$\theta_{\tilde{n},m}^{L}$ and $\theta_{\tilde{n},m}^{H}$.

(A) Due to Eq. (\ref{eq:periodicity}), $\theta_{\tilde{n},m}^{H}$
is equivalently expressed as (in terms of $r''=r'+M/2$) 
\begin{equation}
\theta_{\tilde{n},m}^{H}=\frac{1}{M}\sum_{\left(r'',s\right)\in\mathcal{M}\backslash\left(\mathcal{M}_{c}\cup\left\{ \left(-\frac{M}{2},0\right)\right\} \right)}\Theta_{r'',s}e^{-2\pi i\frac{2r''\tilde{n}+sm}{M}},\label{eq:theta-high2}
\end{equation}
which makes it clear that $\theta_{\tilde{n},m}^{H}$ is a high-pass
filtered map of $\theta_{\tilde{n},m}$. Being high-pass filtered,
we expect elements of $\theta_{\tilde{n},m}^{H}$ to be generally
smaller than the low-pass filtered elements $\theta_{\tilde{n},m}^{L}$
for most natural images (See Sec. \ref{sec:specimen_damage}).

(B) Equations (\ref{eq:conjugate property}), (\ref{eq:theta-low})
and (\ref{eq:theta-high2}) tell us that both the objects $\theta_{\tilde{n},m}^{L},\,\theta_{\tilde{n},m}^{H}$
are approximately real. (It is approximate because, for example, $r'=-M/4$
terms in Eq. (\ref{eq:theta-low}) contribute an imaginary part. The
influence is small for a large $M$.)

(C) Some further equivalent expressions, which is useful for deriving
equations at Step $8$, are 
\[
\theta_{\tilde{n},m}^{L}=\frac{1}{M}\sum_{\left(r,s\right)\in\mathcal{M}_{s}\backslash\mathcal{C}_{s}}\Theta_{r+\frac{M}{4},s}e^{-2\pi i\frac{2\left(r+\frac{M}{4}\right)\tilde{n}+sm}{M}}
\]
\begin{equation}
=\frac{1}{M}\sum_{\left(r,s\right)\in\mathcal{M}_{a}\backslash\mathcal{C}_{a}}\Theta_{r-\frac{M}{4},s}e^{-2\pi i\frac{2\left(r-\frac{M}{4}\right)\tilde{n}+sm}{M}},
\end{equation}
\[
\theta_{\tilde{n},m}^{H}=\frac{1}{M}\sum_{\left(r,s\right)\in\mathcal{M}_{s}\backslash\mathcal{C}_{s}}\Theta_{r-\frac{M}{4},s}e^{-2\pi i\frac{2\left(r+\frac{M}{4}\right)\tilde{n}+sm}{M}}
\]
\begin{equation}
=\frac{1}{M}\sum_{\left(r,s\right)\in\mathcal{M}_{a}\backslash\mathcal{C}_{a}}\Theta_{r+\frac{M}{4},s}e^{-2\pi i\frac{2\left(r-\frac{M}{4}\right)\tilde{n}+sm}{M}}.
\end{equation}
Note that Eq. (\ref{eq:periodicity}), which is due to the discrete
nature of QFT that comes with the use of a quantum computer, is crucial
in deriving some of the above results, including the real-valuedness. 

We are now ready to discuss Steps $7$ and $8$. In Step $7$ we measure
Q2 with respect to basis $\left\{ |\overline{0}\rangle,\,|\overline{1}\rangle\right\} $.
If the result is $|\overline{0}\rangle$, i.e. $c=0$, we are in $\mathcal{M}_{s}$;
and likewise the result $|\overline{1}\rangle$ and $c=1$ means we
are in $\mathcal{M}_{a}$. In Step $8$, we apply split inverse QFT
to Eq. (\ref{eq:step5}), which has been slightly modified in Step
$6$. \begin{widetext} In the case $c=0$, we apply Eq. (\ref{eq:split_iFT1})
to obtain 
\begin{equation}
\frac{1}{M}\sum_{\left(\tilde{n},m\right)\in\mathcal{M}_{c}}|\tilde{n},m\rangle\{\left(1-\alpha\overline{\theta}\right)|\overline{s}\rangle+i\left(\alpha+\overline{\theta}\right)|\overline{a}\rangle+\left[\theta_{\tilde{n},m}^{L}\left(|\overline{s}\rangle+i\alpha|\overline{a}\rangle\right)+\theta_{\tilde{n},m}^{H}\left(|\overline{a}\rangle+i\alpha|\overline{s}\rangle\right)\right]\},\label{eq:r_negative_ifft}
\end{equation}
where the qubit state is for Q1. Likewise, if $c=1$, we perform Eq.
(\ref{eq:split_iFT2}) to obtain 
\begin{equation}
\frac{1}{M}\sum_{\left(\tilde{n},m\right)\in\mathcal{M}_{c}}|\tilde{n},m\rangle\{\left(1-\alpha\overline{\theta}\right)|\overline{a}\rangle+i\left(\alpha+\overline{\theta}\right)|\overline{s}\rangle+\left[\theta_{\tilde{n},m}^{L}\left(|\overline{a}\rangle+i\alpha|\overline{s}\rangle\right)+\theta_{\tilde{n},m}^{H}\left(|\overline{s}\rangle+i\alpha|\overline{a}\rangle\right)\right]\}.\label{eq:r_positive_ifft}
\end{equation}
\end{widetext} Figure 1 (c) shows these states in combination.

Next, the measurement of $\tilde{n},m$ in Step $9$ yields $\hat{n},\,\hat{m}$.
One readily obtains the resultant Q1 state by inspecting Eqs. (\ref{eq:r_negative_ifft})
and (\ref{eq:r_positive_ifft}). Since Step $10$ applies an operation
$|\overline{s}\rangle\Leftrightarrow|\overline{a}\rangle$ to Q1 if
and only if $c=1$, we obtain

\[
|\overline{\psi}\rangle=\left(1-\alpha\overline{\theta}+\theta_{\hat{n},\hat{m}}^{L}+i\alpha\theta_{\hat{n},\hat{m}}^{H}\right)|\overline{s}\rangle
\]
\begin{equation}
+i\left(\alpha+\overline{\theta}+\alpha\theta_{\hat{n},\hat{m}}^{L}-i\theta_{\hat{n},\hat{m}}^{H}\right)|\overline{a}\rangle.
\end{equation}
If $\alpha,\,\theta_{\hat{n},\hat{m}}^{L},\,\theta_{\hat{n},\hat{m}}^{H},\,\overline{\theta}$
are small, we obtain 
\begin{equation}
|\overline{\psi}\rangle=\left(1+\theta_{\hat{n},\hat{m}}^{L}\right)|\overline{s}\rangle+i\left(\alpha+\overline{\theta}-i\theta_{\hat{n},\hat{m}}^{H}\right)|\overline{a}\rangle
\end{equation}
by neglecting second order terms. Further neglecting $\theta_{\hat{n},\hat{m}}^{L},\,\theta_{\hat{n},\hat{m}}^{H}$,
we obtain 
\begin{equation}
|\overline{\psi}\rangle=|\overline{s}\rangle+i\left(\alpha+\overline{\theta}\right)|\overline{a}\rangle.
\end{equation}
Since the entire argument goes through with any small $\alpha$, steps
$1\sim11$ results in $|\overline{\psi}\rangle=|\overline{s}\rangle+ik\overline{\theta}|\overline{a}\rangle$
as desired.

Inclusion of errors $\theta_{\hat{n},\hat{m}}^{L},\,\theta_{\hat{n},\hat{m}}^{H}$,
which vary randomly for each round of $k$ repetitions, have small,
second-order effect in Step $12$. A rough reasoning has been given,
following Eq. (\ref{eq:qubit_state_non_ideal}) in Sec. \ref{subsec:High-level-ideas}.
We defer more complete discussion to Sec. \ref{subsec:Multiple-inelastic-scattering},\ref{subsec:Estimating-the-final},
where an essentially identical situation arises in the setting of
handling inelastic scattering.

\section{Neutralization of inelastic scattering\label{sec:Neutralization-of-inelastic}}

Next, we consider how to deal with inelastic scattering. First, we
sketch the procedure that we later explain in detail in the subsections
below. We only aim to protect a round of measurement here, and do
not try to acquire data from inelastic scattering events. Suppose
that the Q1 state is $|\overline{s}\rangle+i\alpha|\overline{a}\rangle$
before Step 2 and later inelastic scattering occurred in Step $4$.
The state of the entire system is 
\begin{equation}
|s'\overline{s}\rangle+|a'\overline{a}\rangle+i\alpha\left(|s'\overline{a}\rangle+|a'\overline{s}\rangle\right),
\end{equation}
where $|s'\rangle,\,|a'\rangle$ are inelastically scattered states
from $|s\rangle,\,|a\rangle$, respectively. Complete inelastic scattering
neutralization (ISN) would be possible if we could measure the electron
in $|s'\rangle$ \emph{or} $|a'\rangle$. However, the incident beams
$|s\rangle$ and $|a\rangle$ inevitably mix after scattering because
of the overlapping tails of the spread wave functions in the far field.
Our goal is to detect the electron in a state that is as close to
$|s\rangle$ or $|a\rangle$ as possible before resuming the procedure
at Step $2$ with a new electron. Before proceeding, note that it
is \emph{in principle} possible to know the occurrence of inelastic
scattering while preserving the scattered electron state, for example
by measuring the time of flight. 

To meet the goal mentioned above, we mostly follow the same steps
mentioned in Sec. \ref{sec:The-measurement-procedure}. We only replace
Step $6$ with a \emph{randomization step}:

Step $\tilde{6}$: Let $\Xi_{n,m}$ be a set of real numbers that
satisfy $\Xi_{n+M,m}=\Xi_{n,m+M}=\Xi_{n,m}$ and $\Xi_{M/4+n,m}=-\Xi_{M/4-n,-m}$,
but are randomly chosen from $[0,2\pi)$ otherwise. Apply a phase
shift operation $|n,m\rangle\Rightarrow e^{i\Xi_{n,m}}|n,m\rangle$.

This step randomly shifts each SF component in the map $|\tilde{n},m\rangle$
obtained in Step $8$. Specifically, if we get $c=0$ in Step $7$,
by the end of Step $8$ we obtain 
\[
|\Psi'\rangle=\frac{1}{\sqrt{2}}\left\{ |s_{4}\overline{s}\rangle+|a_{4}\overline{a}\rangle+i\alpha\left(|s_{4}\overline{a}\rangle+|a_{4}\overline{s}\rangle\right)\right\} ,
\]
where 
\begin{equation}
|s_{4}\rangle=\frac{i}{M}\sum_{\left(\tilde{n},m\right)\in\mathcal{M}_{c}}h_{\tilde{n},m}^{L}|\tilde{n},m\rangle\label{eq:state_s4}
\end{equation}
and 
\begin{equation}
|a_{4}\rangle=\frac{i}{M}\sum_{\left(\tilde{n},m\right)\in\mathcal{M}_{c}}h_{\tilde{n},m}^{H}|\tilde{n},m\rangle\label{eq:state_a4}
\end{equation}
are a processed version of inelastically scattered states to be discussed
later. (The states $|s_{4}\rangle$ and $|a_{4}\rangle$ are swapped
if $c=1$.) We will show that $h_{\tilde{n},m}^{L}$ and $h_{\tilde{n},m}^{H}$
are real. Hence we have \begin{widetext} 
\begin{equation}
|\Psi'\rangle=\frac{i}{\sqrt{2}M}\sum_{\left(\tilde{n},m\right)\in\mathcal{M}_{c}}|\tilde{n},m\rangle\left\{ h_{\tilde{n},m}^{L}\left(|\overline{s}\rangle+i\alpha|\overline{a}\rangle\right)+h_{\tilde{n},m}^{H}\left(|\overline{a}\rangle+i\alpha|\overline{s}\rangle\right)\right\} .
\end{equation}
\end{widetext}Then, in Step $9$ we measure the electron state $|\tilde{n},m\rangle$.
We are then left with a Q1 state 
\begin{equation}
h_{\tilde{n},m}^{L}\left(|\overline{s}\rangle+i\alpha|\overline{a}\rangle\right)+h_{\tilde{n},m}^{H}\left(|\overline{a}\rangle+i\alpha|\overline{s}\rangle\right),\label{eq:Q1_state_after_split_invFT}
\end{equation}
where $\tilde{n},m$ are the measurement outcomes. We will show that
$\left|h_{\tilde{n},m}^{L}\right|\gg\left|h_{\tilde{n},m}^{H}\right|$
is probable. Hence the state in Eq. (\ref{eq:Q1_state_after_split_invFT})
is approximately $|\overline{s}\rangle+i\alpha|\overline{a}\rangle$
(or $|\overline{a}\rangle+i\alpha|\overline{s}\rangle$ if $c=1$,
which can be converted to $|\overline{s}\rangle+i\alpha|\overline{a}\rangle$
by an operation $|\overline{s}\rangle\Longleftrightarrow|\overline{a}\rangle$).
Thus, we are able to mostly recover the original Q1 state $|\overline{s}\rangle+i\alpha|\overline{a}\rangle$,
successfully neutralizing the adverse effect of inelastic scattering.
Further analysis reveals that the Q1 state has the form $|\overline{s}\rangle+\left(i\alpha+\eta\right)|\overline{a}\rangle$,
where $\eta\in\mathbb{R}$ is generally small deviation from the ideal,
whose magnitude will be estimated to evaluate the performance of our
method (See Fig. 2 (c)). In the rest of this Section, we discuss these
procedures in detail.

\subsection{Preliminary remarks\label{subsec:Preliminary-remarks}}

In general, upon inelastic scattering, the specimen is excited to
multiple states and hence the scattered probe electron and the specimen
get entangled. This gives a mixed probe electron state. However, the
mixed nature of scattered electron state does not play a significant
role because the scattered probe electron state only weakly depends
on the final state of the specimen as we describe in subsection C
of Appendix C. Hence, for simplicity we assume that the state of the
specimen goes from the ``ground'' state $|g\rangle$ (More precisely,
this is merely an initial state but we will call it the ground state
hereafter.) to a particular excited state $|e\rangle$. Let the energy
difference between the states $|g\rangle,|e\rangle$ be $E$.

In analysing ISN, we need to know the wavefunction of inelastically
scattered electrons. In Appendix C, we briefly review theory of inelastic
scattering \cite{Egerton_textbook,Messiah_qm} and obtain the functional
form of such a wavefunction under the assumption that we are in the
dipole region. Here we describe only the result. Recall that the scattering
vector is $\mathbf{q}=\mathbf{k}_{f}-\mathbf{k}_{i}$. We define $\Delta k\equiv k_{i}-k_{f}$,
which is a function of the energy loss $E$, whose typical value for
exciting a plasmon is $\approx20\,\mathrm{eV}$. Let the scattering
angle be $\theta$. Define $\theta_{E}\equiv\frac{\Delta k}{k_{i}}\approx\frac{E}{2E_{K}}$,
where $E_{K}$ is the energy of incident electrons, which we assume
to be $300\,\mathrm{keV}$. Let $\theta_{c}=\sqrt{\frac{2\theta_{E}}{\gamma}}$
be the Bethe ridge angle. For the values of $E_{K}$ and $E$ mentioned
above, the angles $\theta_{E}$ and $\theta_{c}$ have values $41\,\mu\mathrm{rad}$
and $7.2\,\mathrm{mrad}$, respectively. Under a further reasonable
assumption of ``achirality'' (See Appendix C) we obtain the form
of wavefunction as

\begin{equation}
\Psi_{0}\left(\mathbf{q}\right)=\begin{cases}
\frac{1}{\sqrt{\theta^{2}+\theta_{E}^{2}}}\left(\frac{\mathbf{q}}{q}\right)\cdot\mathbf{a} & \theta<\theta_{c}\\
0 & \theta>\theta_{c}
\end{cases}\label{eq:Psi1}
\end{equation}
where $\mathbf{a}$ is the unknown direction of the dipole involved
in inelastic scattering in the dipole region. The magnitude of $\mathbf{a}$
is unimportant because the wavefunction needs to be normalized anyway.
Define $\psi_{0}\left(\mathbf{r}\right)=\mathcal{F}_{C}^{-1}\left\{ \Psi_{0}\left(\mathbf{k}\right)\right\} $,
which is the real-space wavefunction right after inelastic scattering.

Since we can not know the location of inelastic scattering $\mathbf{r}_{0}$,
we should use a slightly generalized form of wavefunction (See Appendix
C) 
\begin{equation}
\Psi_{1}\left(\mathbf{q}\right)=e^{-i\mathbf{q}\cdot\mathbf{r}_{0}}\Psi_{0}\left(\mathbf{q}\right),
\end{equation}
for which the following holds: 
\begin{equation}
\psi_{1}\left(\mathbf{r}\right)=\mathcal{F}_{C}^{-1}\left\{ \Psi_{1}\left(\mathbf{k}\right)\right\} =\psi_{0}\left(\mathbf{r}-\mathbf{r}_{0}\right).
\end{equation}

\subsection{An array of inelastically scattered focused beams and their discrete
Fourier transform $E_{n,m}$}

For simplicity, first consider a single incident beam focused at $\mathbf{r}_{1}$.
The wavefunction in the $xy$-plane is $\psi\left(\mathbf{r}\right)=\delta^{2}\left(\mathbf{r}-\mathbf{r}_{1}\right)$.
Fourier transforming, we obtain the far-field wavefunction in the
upstream of electron optics as 
\begin{equation}
\Psi\left(\mathbf{k}_{i}\right)=\mathcal{F}_{C}\left\{ \psi\left(\mathbf{r}\right)\right\} =e^{-i\mathbf{k}_{i}\cdot\mathbf{r}_{1}}.
\end{equation}
We assume that inelastic excitation is insensitive to the angular
variation of incident plane wave measured in $\approx10^{-3}\,\mathrm{rad}$.
The scattered waves are superposition of $\Psi_{1}\left(\mathbf{q}\right)$
because the incident wave is a superposition of plane waves. Using
the principle of superposition, after inelastic scattering we obtain,
in the far field 
\begin{equation}
\Psi_{2}\left(\mathbf{q}\right)=\int\Psi_{1}\left(\mathbf{k}_{f}-\mathbf{k}_{i}\right)e^{-i\mathbf{k}_{i}\cdot\mathbf{r}_{1}}\frac{d^{2}\mathbf{k}_{i}}{\left(2\pi\right)^{2}},
\end{equation}
which has a form of convolution. Thus, defining $\psi_{2}\left(\mathbf{r}\right)=\mathcal{F}_{C}^{-1}\left\{ \Psi_{2}\left(\mathbf{q}\right)\right\} $,
we have 
\begin{equation}
\psi_{2}\left(\mathbf{r}\right)=\psi_{1}\left(\mathbf{r}\right)\delta^{2}\left(\mathbf{r}-\mathbf{r}_{1}\right)=\psi_{0}\left(\mathbf{r}-\mathbf{r}_{0}\right)\delta^{2}\left(\mathbf{r}-\mathbf{r}_{1}\right).
\end{equation}
The physical meaning of this is clear because this is the state right
after a scattering event centered at $\mathbf{r}_{0}$, when the incident
electron wave is focused at $\mathbf{r}_{1}$.

Having found the state right after inelastic scattering for a single
focused incident electron beam, we use the principle of superposition
to straightforwardly generalize it to the case of an array of focused
incident beams. (``An array of focused beams'' may be a misnomer,
since we only have a single electron.) Consider the array of focused
incident electron beams, i.e. 
\begin{equation}
|s\rangle=\frac{1}{M}\sum_{\left(n,m\right)\in\mathcal{M}}e^{i\frac{\pi}{2}n}|n,m\rangle.\label{eq:swave}
\end{equation}
This electron wave, comprising $M^{2}$ focused beams, passes the
specimen, and experiences inelastic scattering. Here we ignore small
phase shift due to simultaneous elastic scattering, which is a result
of only a single passage of the electron wave. The resultant state
is 
\begin{equation}
|s_{2}\rangle=\frac{1}{M}\sum_{\left(n,m\right)\in\mathcal{M}}\psi_{0}\left(\mathbf{r}_{n,m}-\mathbf{r}_{0}\right)e^{i\frac{\pi}{2}n}|n,m\rangle,\label{eq:swave_scattered}
\end{equation}
where $\mathbf{r}_{n,m}=n\sigma\hat{\mathbf{i}}+m\sigma\hat{\mathbf{j}}$.
We assume that $M$ is sufficiently large that we can ignore the possibility
of inelastic scattering occurring at near the edge of the $M\times M$
array. Also for simplicity, write 
\begin{equation}
e_{n,m}=\psi_{0}\left(\mathbf{r}_{n,m}-\mathbf{r}_{0}\right)\label{eq:discrete_psi0}
\end{equation}
to describe a discretized version of the wavefunction right after
inelastic scattering. Next, we apply QFT defined in Eqs. (\ref{eq:DFT},
\ref{eq:q_state_before_FT}, \ref{eq:Fourier_transformed_q_state})
to Eq. (\ref{eq:swave_scattered}). We obtain as a result 
\begin{equation}
|s_{3}\rangle=\frac{1}{M}\sum_{\left(n,m\right)\in\mathcal{M}}E_{n,m}^{\left(s\right)}|n,m\rangle,
\end{equation}
where 
\begin{equation}
E_{n,m}^{\left(s\right)}=\frac{1}{M}\sum_{\left(r,s\right)\in\mathcal{M}}e_{r,s}e^{2\pi i\frac{\left(n+\frac{M}{4}\right)r+ms}{M}}.\label{eq:E_nm^s_definition}
\end{equation}
Following exactly the same argument, but for the other state $|a\rangle,$we
obtain 
\begin{equation}
E_{n,m}^{\left(a\right)}=\frac{1}{M}\sum_{\left(r,s\right)\in\mathcal{M}}e_{r,s}e^{2\pi i\frac{\left(n-\frac{M}{4}\right)r+ms}{M}}.\label{eq:E_nm^a_definition}
\end{equation}
From these relations, we find $E_{n+M,m}=E_{n,m}$ for both $E_{n,m}^{\left(s\right)}$
and $E_{n,m}^{\left(a\right)}$; and $E_{n+\frac{M}{2},m}^{\left(s\right)}=E_{n,m}^{\left(a\right)}$.

\subsection{Symmetry relations for $E_{n,m}^{\left(s\right)}$ and $E_{n,m}^{\left(a\right)}$ }

The quantities $E_{n,m}^{\left(s\right)}$ and $E_{n,m}^{\left(a\right)}$
satisfy certain relations that we call \emph{symmetry relations}.
The intuition is the following. The set of amplitudes $e_{n,m}$,
which reflects the wavefunction right after inelastic scattering turns
out to be pure imaginary. The quantities $E_{n,m}^{\left(s\right)}$
and $E_{n,m}^{\left(a\right)}$ are modified versions of DFT of $e_{r,s}$.
Hence $E_{n,m}^{\left(s\right)}$ and $E_{n,m}^{\left(a\right)}$,
which are akin to far-field wavefunctions, must satisfy certain relations
involving complex conjugation, similarly to the far-field wavefunction
of a transmitted wave through a pure weak phase object. In addition,
unlike coutinuous FT found in electron optics, we have certain additional
symmetries because of the periodic nature of DFT.\begin{widetext}

\begin{figure}[t]
\includegraphics[scale=0.35]{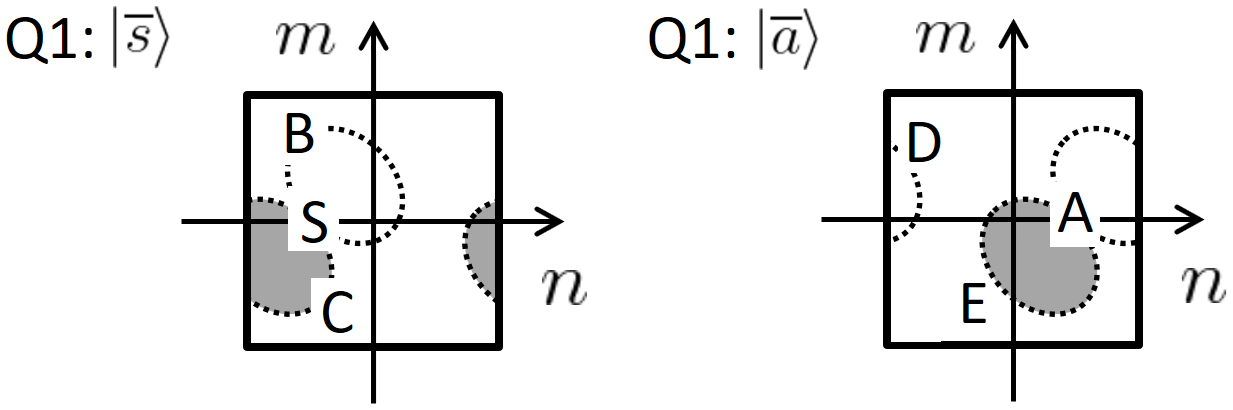}(a)

\includegraphics[scale=0.35]{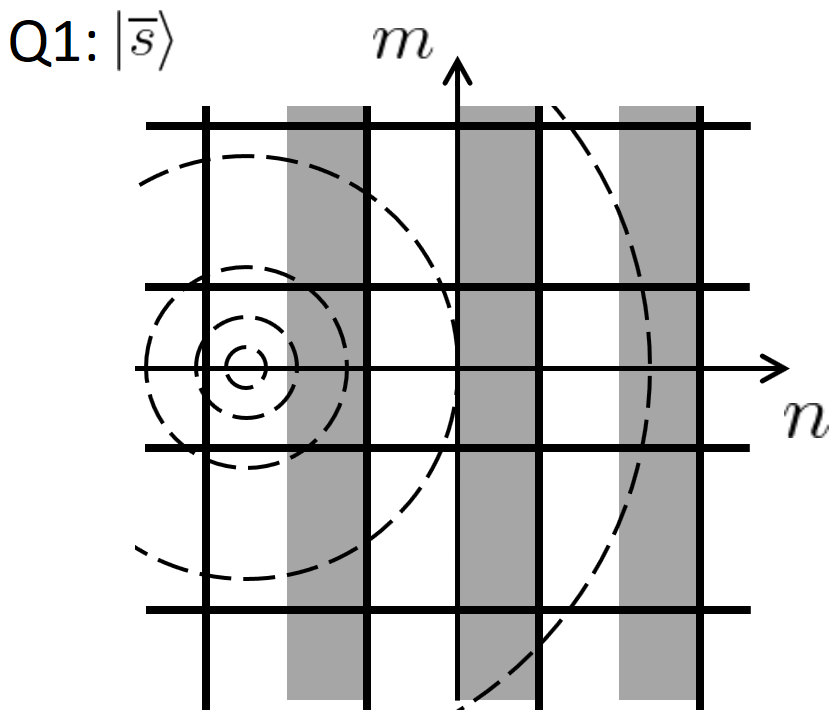}(b)\includegraphics[scale=0.35]{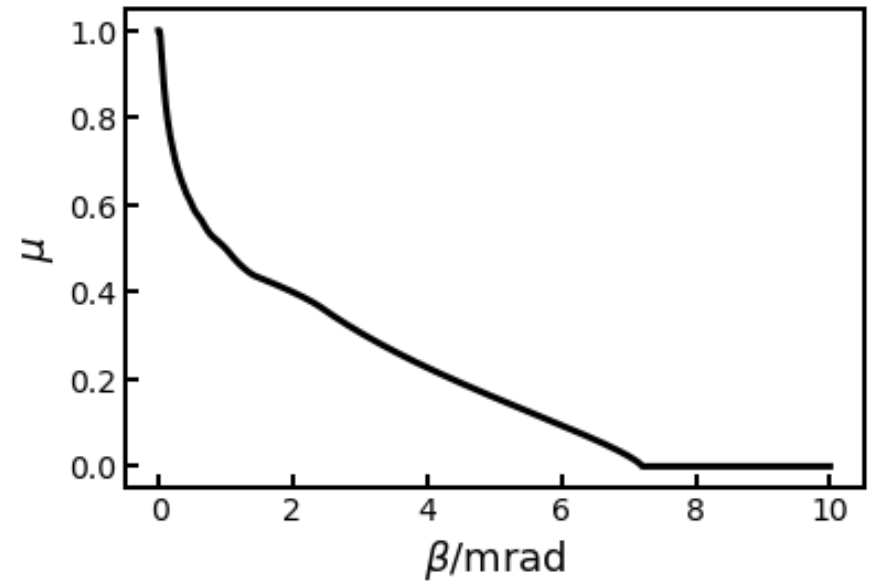}(c)

\caption{Inelastic scattering neutralization (ISN). (a) A state appearing in
inelastic scattering neutralization. We schematically show the inelastically
scattered wavefunction $\Psi_{0}\left(\mathbf{k}\right)$ in the far
field, which is an odd function, with unknown dipole orientation $\mathbf{a}$
(here, $\mathbf{a}$ points towards the upper-right direction). The
white and gray regions indicates positive and negative regions of
$\Psi_{0}\left(\mathbf{k}\right)$, although such a characterization
is oversimplified because the wavefront should also have the phase
factor $e^{-i\mathbf{q}\cdot\mathbf{r}_{0}}$ that represents tilt.
As is the case with elastic scattering, the amplitudes at B and C,
or D and E, are the negative of complex conjugate (NCC) to each other.
This helps to generate $\eta\in\mathbb{R}$. (b) An averaged intensity
map of inelastically scattered electron waves in the far field. This
is one branch of the far field wherein Q1 state is $|\overline{s}\rangle$.
Although the far field has a finite area (call it a \emph{cell}) with
the periodic boundary condition (PBC), here it is shown in a expanded
way. The \emph{intensity} of inelastically scattered electron wave,
which is averaged over all dipole orientations $\mathbf{a}$, is shown
in dashed circles. Although this intensity pattern is repeated due
to PBC, only the one originating from the point S of a single cell
is shown for clarity. The white regions correspond to the set $\mathcal{S}$
described in the text, while the gray regions correspond to $\mathcal{A}$.
If one focuses on an area of either $\mathcal{S}$ or $\mathcal{A}$
type in a particular cell, then one should see that the wave intensity
in that area comes from waves originating from many different cells.
Subsequently, waves in $\mathcal{M}_{s}$ and $\mathcal{M}_{a}$ regions,
respectively, are mixed up within these regions in the cell during
the split inverse FT. Equations (\ref{eq:S_integ_S8}, \ref{eq:A_integ_S8})
express incoherent addition of these waves. (c) This plot shows how
$\mu\approx\left|\eta\right|$, which quantifies nonideality of ISN,
changes with $\beta=\frac{\lambda}{2\sigma}$, where $\sigma$ is
the resolution of interest. Things get better at \emph{higher} resolution.}
\end{figure}

\end{widetext}

Indeed, $e_{n,m}=\psi_{0}\left(\mathbf{r}_{n,m}-\mathbf{r}_{0}\right)$
is pure imaginary. Equation (\ref{eq:Psi1}), combined with 
\begin{equation}
\psi_{0}\left(\mathbf{r}\right)=\mathcal{F}_{C}^{-1}\left\{ \Psi_{0}\left(\mathbf{k}\right)\right\} =\int\Psi_{0}\left(\mathbf{k}\right)e^{i\mathbf{k}\cdot\mathbf{r}}\frac{d^{2}\mathbf{k}}{\left(2\pi\right)^{2}}
\end{equation}
shows that 
\[
\psi_{0}\left(\mathbf{r}\right)^{*}=\int\Psi_{0}\left(\mathbf{k}\right)e^{-i\mathbf{k}\cdot\mathbf{r}}\frac{d^{2}\mathbf{k}}{\left(2\pi\right)^{2}}=\int\Psi_{0}\left(-\mathbf{k}\right)e^{i\mathbf{k}\cdot\mathbf{r}}\frac{d^{2}\mathbf{k}}{\left(2\pi\right)^{2}}
\]
\begin{equation}
=-\int\Psi_{0}\left(\mathbf{k}\right)e^{i\mathbf{k}\cdot\mathbf{r}}\frac{d^{2}\mathbf{k}}{\left(2\pi\right)^{2}}=-\psi_{0}\left(\mathbf{r}\right).
\end{equation}
Hence we obtain $e_{n,m}^{*}=-e_{n,m}$.

Now we exhibit the symmetry relations. Since exactly the same relations
hold both for $E_{n,m}^{\left(s\right)}$ and $E_{n,m}^{\left(a\right)}$,
we write these generically as $E_{n,m}$. These are 
\begin{equation}
E_{\frac{M}{4}+a,b}=-E_{\frac{M}{4}-a,-b}^{*},\label{eq:symmetry1}
\end{equation}
and 
\begin{equation}
E_{-\frac{M}{4}+a,b}=-E_{-\frac{M}{4}-a,-b}^{*}.\label{eq:symmetry2}
\end{equation}
These relations are straightforwardly obtained from Eqs. (\ref{eq:E_nm^s_definition},
\ref{eq:E_nm^a_definition}) and the relation $e_{n,m}^{*}=-e_{n,m}$.
Equations (\ref{eq:symmetry1}, \ref{eq:symmetry2}) are crucial for
keeping certain quantities real in later steps.

Figure 2 (a) shows the state of the system at this point. (See also
the explanations of Fig. 1 for information about how to view the figure.)
The symmetry relations make the quantum amplitudes at B and C, or
those of D and E, NCC of each other. The dumbbell structure symbolically
illustrates the ``dipole'' nature of $\Psi_{0}\left(\mathbf{k}\right)$.

\subsection{The randomization step\label{subsec:randomization_step}}

To strengthen later assumptions described in Sec. \ref{subsec:eta_estimation_S8},
here we ``randomize'' the coefficients $E_{n,m}^{\left(s\right)}$
and $E_{n,m}^{\left(a\right)}$. Again, we denote them collectively
as $E_{n,m}$ because it can equally be taken as $E_{n,m}^{\left(s\right)}$
or $E_{n,m}^{\left(a\right)}$. This is the \emph{randomization step}.
Let a set of real values be $\Xi_{n,m}$, which satisfy 
\begin{equation}
\Xi_{n+M,m}=\Xi_{n,m+M}=\Xi_{n,m},\label{eq:Xi_constraint1_S8}
\end{equation}
\begin{equation}
\Xi_{\frac{M}{4}+n,m}=-\Xi_{\frac{M}{4}-n,-m}.\label{eq:Xi_constraint2_S8}
\end{equation}
When not constrained by Eqs. (\ref{eq:Xi_constraint1_S8}, \ref{eq:Xi_constraint2_S8}),
$\Xi_{n,m}$ are chosen at random from $[0,\,2\pi)$. From these relations,
one finds $\Xi_{\frac{M}{4}+n,m}=-\Xi_{-\frac{3M}{4}-n,-m}$, and
then obtain, by replacing $n$ with $n-\frac{M}{2}$, 
\begin{equation}
\Xi_{-\frac{M}{4}+n,m}=-\Xi_{-\frac{M}{4}-n,-m}.
\end{equation}
One can verify that the replacement of $E_{n,m}$ with $E_{n,m}e^{i\Xi_{n,m}}$,
which constitutes the randomization step, does not violate the symmetry
relations.

\subsection{The split inverse Fourier transform}

We show that application of split inverse QFT on $E_{n,m}$, after
the randomization step, yields $g_{n,m}$ that is purely imaginary,
because of the symmetry relations. By Eq. (\ref{eq:split_iFT1}),
we have 
\[
g_{\tilde{n},m}^{L}=\frac{\sqrt{2}}{M}\sum_{\left(r,s\right)\in\mathcal{M}_{s}}E_{r,s}^{\left(s\right)}e^{-2\pi i\frac{2\left(r+\frac{M}{4}\right)\tilde{n}+sm}{M}}
\]
\begin{equation}
=\frac{\sqrt{2}}{M}\sum_{\left(r,s\right)\in\mathcal{M}_{c}}E_{r-\frac{M}{4},s}^{\left(s\right)}e^{-2\pi i\frac{2r\tilde{n}+sm}{M}},\label{eq:Es_to_gL_isFT}
\end{equation}
and by Eq. (\ref{eq:split_iFT2}) we obtain 
\[
g_{\tilde{n},m}^{H}=\frac{\sqrt{2}}{M}\sum_{\left(r,s\right)\in\mathcal{M}_{a}}E_{r,s}^{\left(s\right)}e^{-2\pi i\frac{2\left(r-\frac{M}{4}\right)\tilde{n}+sm}{M}}
\]
\begin{equation}
=\frac{\sqrt{2}}{M}\sum_{\left(r,s\right)\in\mathcal{M}_{c}}E_{r+\frac{M}{4},s}^{\left(s\right)}e^{-2\pi i\frac{2r\tilde{n}+sm}{M}}.\label{eq:Es_to_gH_isFT}
\end{equation}
Superscripts L, H denote ``low-pass filtered'' and ``high-pass
filtered'', respectively, in a similar way with the elastic scattering
case. Replacement of $E_{r,s}^{\left(s\right)}$ with $E_{r,s}^{\left(a\right)}$
results in swapping of $g_{\tilde{n},m}^{H}$ and $g_{\tilde{n},m}^{L}$,
as in 
\begin{equation}
g_{\tilde{n},m}^{H}=\frac{\sqrt{2}}{M}\sum_{\left(r,s\right)\in\mathcal{M}_{s}}E_{r,s}^{\left(a\right)}e^{-2\pi i\frac{2\left(r+\frac{M}{4}\right)\tilde{n}+sm}{M}}\label{eq:Ea_to_gH_isFT}
\end{equation}
and 
\begin{equation}
g_{\tilde{n},m}^{L}=\frac{\sqrt{2}}{M}\sum_{\left(r,s\right)\in\mathcal{M}_{a}}E_{r,s}^{\left(a\right)}e^{-2\pi i\frac{2\left(r-\frac{M}{4}\right)\tilde{n}+sm}{M}}.\label{eq:Ea_to_gL_isFT}
\end{equation}
Next, we take complex conjugation of $g_{\tilde{n},m}^{L}$: 
\[
\left(g_{\tilde{n},m}^{L}\right)^{*}=\frac{\sqrt{2}}{M}\sum_{\left(r,s\right)\in\mathcal{M}_{c}}E_{r-\frac{M}{4},s}^{*}e^{2\pi i\frac{2r\tilde{n}+sm}{M}}
\]
\[
=-\frac{\sqrt{2}}{M}\sum_{\left(r,s\right)\in\mathcal{M}_{c}}E_{-\frac{M}{4}-r,-s}e^{2\pi i\frac{2r\tilde{n}+sm}{M}}
\]
\begin{equation}
=-\frac{\sqrt{2}}{M}\sum_{\left(r',s'\right)\in\mathcal{M}_{c}}E_{-\frac{M}{4}+r',s'}e^{-2\pi i\frac{2r'\tilde{n}+s'm}{M}}=-g_{\tilde{n},m}^{L},
\end{equation}
where $\left(r,s\right)=\left(-r',-s'\right)$. Following similar
steps, we obtain $\left(g_{\tilde{n},m}^{H}\right)^{*}=-g_{\tilde{n},m}^{H}$.
Hence, for both low-pass and high-pass filtered versions, $g_{\tilde{n},m}$
is purely imaginary. Define $h_{\tilde{n},m}\in\mathbb{R}$ that satisfy
$g_{\tilde{n},m}=ih_{\tilde{n},m}$.

Thus, we obtain Eqs. (\ref{eq:state_s4}, \ref{eq:state_a4}) as resultant
states after performing split inverse FT if $c=0$. We obtain 
\begin{equation}
|a_{4}\rangle=\frac{i}{M}\sum_{\left(\tilde{n},m\right)\in\mathcal{M}_{c}}h_{\tilde{n},m}^{L}|\tilde{n},m\rangle
\end{equation}
and 
\begin{equation}
|s_{4}\rangle=\frac{i}{M}\sum_{\left(\tilde{n},m\right)\in\mathcal{M}_{c}}h_{\tilde{n},m}^{H}|\tilde{n},m\rangle
\end{equation}
if $c=1$.

\subsection{Estimating $\left|\eta\right|$ that quantifies nonideality \label{subsec:eta_estimation_S8}}

Next, we estimate the parameter $\eta$ that appeared, following Eq.
(\ref{eq:Q1_state_after_split_invFT}). We begin with a heuristic
discussion. Consider Eq. (\ref{eq:E_nm^s_definition}) givng $E_{n,m}^{\left(s\right)}$.
This is FT of $e_{n,m}$, whose center is placed at $\left(n,m\right)=\left(-\frac{M}{4},0\right)$.
On the other hand, $e_{n,m}=\psi_{0}\left(\mathbf{r}_{n,m}-\mathbf{r}_{0}\right)$
and hence its FT is essentially $\Psi_{0}\left(\mathbf{k}\right)e^{-i\mathbf{k}\cdot\mathbf{r}_{0}}$.
More precisely, the correspondence before the randomization step is
\begin{equation}
E_{n,m}^{\left(s\right)}\approx F\Psi_{1}\left(\mathbf{k}_{-n-\frac{M}{4},-m}\right)\label{eq:E_nm_and_Psi_relation}
\end{equation}
where $\mathbf{k}_{n,m}=k_{\mathrm{min}}\left(n\hat{\mathbf{i}}+m\hat{\mathbf{j}}\right)$
and $F$ is a proportionality constant. The negative signs are a consequence
of the unfortunate difference of convention between continuous and
discrete FT. 

Since $\Psi_{0}\left(\mathbf{k}\right)$ is large only when $\mathbf{k}$
is close to zero, $E_{n,m}^{\left(s\right)}$ clearly is large at
around $n\approx-\frac{M}{4},\,m\approx0$. By a similar argument,
we find that $E_{n,m}^{\left(a\right)}$ is large near $n\approx\frac{M}{4},\,m\approx0$.
Since split inverse FT is performed separately in regions $\mathcal{M}_{s}$
and $\mathcal{M}_{a}$, $\left|h_{\tilde{n},m}^{L}\right|$ tends
to be larger than $\left|h_{\tilde{n},m}^{H}\right|$.

We present estimation of the magnitude of $\left|\eta\right|$, which
we believe is reasonably accurate and conceptually transparent. To
make this problem tractable, first we make a quite reasonable assumption
that $\left|h_{n,m}^{L}\right|>\left|h_{n,m}^{H}\right|$ mostly holds.
Hence we obtain 
\begin{equation}
\left|\eta\right|=\sum_{\left(\tilde{n},m\right)\in\mathcal{M}_{c}}p_{\tilde{n},m}\left|\frac{h_{\tilde{n},m}^{H}}{h_{\tilde{n},m}^{L}}\right|,\label{eq:absolute_eta_value1.5}
\end{equation}
where 
\[
p_{\tilde{n},m}\propto\left|h_{\tilde{n},m}^{L}+i\alpha h_{\tilde{n},m}^{H}\right|^{2}+\left|h_{\tilde{n},m}^{L}+i\alpha h_{\tilde{n},m}^{H}\right|^{2}
\]
\begin{equation}
\approx\left|h_{\tilde{n},m}^{L}\right|^{2}+\left|h_{\tilde{n},m}^{H}\right|^{2}
\end{equation}
Second, note that the randomization step described above made all
pixels ``equal'', in the sense that all spatial frequency components,
which are sinusoidal, are randomly shifted in the real space and hence
there is no special location in the real space. This encourages us
to use an approximation 
\begin{equation}
p_{\tilde{n},m}=\frac{2}{M^{2}}.
\end{equation}
Third, we make a standard approximation that root-mean-square roughly
equals the mean of absolute values, i.e., 
\begin{equation}
\left|\eta\right|\approx\sqrt{\frac{2}{M^{2}}\sum_{\left(\tilde{n},m\right)\in\mathcal{M}_{c}}\left(\frac{h_{\tilde{n},m}^{H}}{h_{\tilde{n},m}^{L}}\right)^{2}}.\label{eq:absolute_eta_value1.8}
\end{equation}

At this point, we make a brief, purely mathematical, digression to
investigate whether the mean of ratios can be replaced by the ratio
of means. More specifically, let $a_{n},\,b_{n}$ be series, where
$n=1,2,\cdots,N$. We assume that $a_{n}\ll b_{n}$ for all $n$.
Hence we write $a_{n}=\varepsilon_{n}b_{n}$, where $\varepsilon_{n}\ll1$.
Write this series $\varepsilon_{n}=\varepsilon+\delta_{n}$, where
$\sum_{n}\delta_{n}=0$. Also write $b_{n}=b+d_{n}$, where $\sum_{n}d_{n}=0$.
Consider the mean of ratios 
\begin{equation}
\frac{1}{N}\sum_{n}\frac{a_{n}}{b_{n}}=\frac{1}{N}\sum_{n}\varepsilon_{n}=\varepsilon+\frac{1}{N}\sum_{n}\delta_{n}=\varepsilon.
\end{equation}
On the other hand, the ratio of means is 
\[
\frac{\frac{1}{N}\sum_{n}a_{n}}{\frac{1}{N}\sum_{n}b_{n}}=\frac{\sum_{n}\varepsilon_{n}b_{n}}{\sum_{n}b_{n}}
\]
\[
=\frac{\varepsilon\sum_{n}b_{n}+\sum_{n}\delta_{n}b_{n}}{\sum_{n}b_{n}}=\varepsilon+\frac{\sum_{n}\delta_{n}b_{n}}{\sum_{n}b_{n}}
\]
\begin{equation}
=\varepsilon+\frac{b\sum_{n}\delta_{n}+\sum_{n}\delta_{n}d_{n}}{\sum_{n}b_{n}}=\varepsilon+\frac{\sum_{n}\delta_{n}d_{n}}{Nb}.\label{eq:ratio_of_means_S8}
\end{equation}
Comparing these, we conclude that the mean of ratio is close to the
ratio of means if two series $\delta_{n},\,d_{n}$ are only weakly
correlated.

Fourth, going back to the main line of reasoning, we assume that the
conditions for the above ``theorem'' are met with $\left(h_{\tilde{n},m}^{H}\right)^{2}$
and $\left(h_{\tilde{n},m}^{L}\right)^{2}$, where the former is identified
with $a_{n}$ while the latter corresponds to $b_{n}$. Then, we may
use what we have just shown to obtain, from Eq. (\ref{eq:absolute_eta_value1.8}),
\begin{equation}
\left|\eta\right|\approx\sqrt{\frac{\sum_{\left(\tilde{n},m\right)\in\mathcal{M}_{c}}\left(h_{\tilde{n},m}^{H}\right)^{2}}{\sum_{\left(\tilde{n},m\right)\in\mathcal{M}_{c}}\left(h_{\tilde{n},m}^{L}\right)^{2}}}.
\end{equation}
Application of Parseval's theorem to Eq. (\ref{eq:Es_to_gL_isFT})
yields 
\begin{equation}
\sum_{\left(\tilde{n},m\right)\in\mathcal{M}_{c}}\left(h_{\tilde{n},m}^{L}\right)^{2}=\sum_{\left(n,m\right)\in\mathcal{M}_{s}}\left|E_{n,m}^{\left(s\right)}\right|^{2}.
\end{equation}
From Eq. (\ref{eq:Ea_to_gH_isFT}), we also obtain 
\begin{equation}
\sum_{\left(\tilde{n},m\right)\in\mathcal{M}_{c}}\left(h_{\tilde{n},m}^{H}\right)^{2}=\sum_{\left(n,m\right)\in\mathcal{M}_{s}}\left|E_{n,m}^{\left(a\right)}\right|^{2}.
\end{equation}
On the other hand, Eq. (\ref{eq:E_nm^s_definition}) says, for all
$\left(r,s\right)\in\mathbb{Z}^{2}$, 
\begin{equation}
E_{n,m}^{\left(s\right)}=E_{n+Mr,m+Ms}^{\left(s\right)}.
\end{equation}
We incorporate this periodicity in Eq. (\ref{eq:E_nm_and_Psi_relation})
to obtain more accurate expression: 
\begin{equation}
E_{n,m}^{\left(s\right)}\approx F\sum_{\left(r,s\right)\in\mathbb{Z}^{2}}\Psi_{1}\left(\mathbf{k}_{-n-M\left(r+\frac{1}{4}\right),-m-Ms}\right)e^{i\Xi_{n,m}}.
\end{equation}
At this point, we make further approximation that the phase factor
in this equation is totally random. The presence of randomization
step described in Sec. \ref{subsec:randomization_step} prompts us
to accept this assumption. We use a mathematical identity 
\begin{equation}
\left|\sum_{k}a_{k}e^{i\theta_{k}}\right|^{2}=\sum_{k}\left|a_{k}\right|^{2}+2\sum_{k\neq l}a_{k}a_{l}\cos\left(\theta_{k}-\theta_{l}\right),\label{eq:math_identity3}
\end{equation}
where $a_{k}\in\mathbb{R}$ and the second term averages to zero when
$\theta_{k}$ are all random. We obtain 
\begin{equation}
\left|E_{n,m}^{\left(s\right)}\right|^{2}\approx F^{2}\left\{ \sum_{\left(r,s\right)\in\mathbb{Z}^{2}}\left|\Psi_{1}\left(\mathbf{k}_{-n-M\left(r+\frac{1}{4}\right),-m-Ms}\right)\right|^{2}\right\} .
\end{equation}
Recall that $\Psi_{0}\left(\mathbf{k}\right)$, mentioned in Eq. (\ref{eq:Psi1}),
includes a particular direction $\mathbf{a}$ of the dipole. This
direction should be regarded random and our analysis should be averaged
over the direction of $\mathbf{a}$. We assume that such averaging
can be done here, rather than at the final stage of computing $\left|\eta\right|$
without introducing significant error. Equation (\ref{eq:Psi1}) can
be expressed as 
\begin{equation}
\Psi_{0}\left(\mathbf{k}\right)=\begin{cases}
\frac{a\cos\xi}{\sqrt{\theta^{2}+\theta_{E}^{2}}} & \theta<\theta_{c}\\
0 & \theta>\theta_{c}
\end{cases},
\end{equation}
where $\xi$ is the angle between $\mathbf{k}$ and $\mathbf{a}$.
Hence 
\begin{equation}
\left|\Psi_{1}\left(\mathbf{k}\right)\right|^{2}=\left|\Psi_{0}\left(\mathbf{k}\right)\right|^{2}=\begin{cases}
\frac{a^{2}\cos^{2}\xi}{\theta^{2}+\theta_{E}^{2}} & \theta<\theta_{c}\\
0 & \theta>\theta_{c}
\end{cases}
\end{equation}
To compute the average of $\cos^{2}\xi$ in $3$ dimensional space,
we take the $z$-axis parallel to $\mathbf{k}$ and move $\frac{\mathbf{a}}{a}$
on the unit sphere $\mathcal{U}$. The average of $\cos^{2}\xi$ is,
using obvious notations of polar coordinates, 
\begin{equation}
\int_{\mathcal{U}}\cos^{2}\xi\frac{dS}{4\pi}=\frac{1}{2}\int_{0}^{\pi}d\xi\cos^{2}\xi\cdot\sin\xi==\frac{1}{3}.
\end{equation}
Henceforth we use the averaged version $\Phi\left(\mathbf{k}\right)$
shown below, instead of $\left|\Psi_{0}\left(\mathbf{k}\right)\right|^{2}$:
\begin{equation}
\Phi\left(\mathbf{k}\right)=\begin{cases}
\frac{a^{2}}{3}\cdot\frac{1}{\theta^{2}+\theta_{E}^{2}} & \theta<\theta_{c}\\
0 & \theta>\theta_{c}
\end{cases},
\end{equation}
where the factor $\frac{a^{2}}{3}$ is not important after all.

Recalling that $\mathbf{k}_{a,b}=k_{\mathrm{min}}\left(a\hat{\mathbf{i}}+b\hat{\mathbf{j}}\right)$,
we approximate a sum by an integral: 
\[
\sum_{\left(n,m\right)\in\mathcal{M}_{s}}\left|E_{n,m}^{\left(s\right)}\right|^{2}k_{\mathrm{min}}^{2}
\]
\begin{equation}
\approx F^{2}\sum_{\left(r,s\right)\in\mathbb{Z}^{2}}\int_{\mathcal{D}}d^{2}\mathbf{k}\Phi\left(\mathbf{k}+k_{\mathrm{max}}\left(r\hat{\mathbf{i}}+s\hat{\mathbf{j}}\right)\right),
\end{equation}
where $\mathcal{D}$ denotes a region \begin{widetext} 
\begin{equation}
\mathcal{D}=\left\{ \left(k_{x,}k_{y}\right)|-\frac{k_{\mathrm{max}}}{4}<k_{x}<\frac{k_{\mathrm{max}}}{4},\,-\frac{k_{\mathrm{max}}}{2}<k_{y}<\frac{k_{\mathrm{max}}}{2}\right\} ,
\end{equation}
\end{widetext}where $\left(k_{x},k_{y}\right)\in\mathbb{R}^{2}$.
Equivalently, one can define a region $\mathcal{S}$, which is a set
of ``stripes'': 
\begin{equation}
\mathcal{S}=\bigcup_{r\in\mathbb{Z}}\left\{ \left(k_{x,}k_{y}\right)|r-\frac{1}{4}<\frac{k_{x}}{k_{\mathrm{max}}}<r+\frac{1}{4}\right\} ,
\end{equation}
to express 
\begin{equation}
\sum_{\left(n,m\right)\in\mathcal{M}_{s}}\left|E_{n,m}^{\left(s\right)}\right|^{2}k_{\mathrm{min}}^{2}\approx F^{2}\int_{\mathcal{\mathcal{S}}}d^{2}\mathbf{k}\Phi\left(\mathbf{k}\right).\label{eq:S_integ_S8}
\end{equation}
A similar argument on $E_{n,m}^{\left(a\right)}$ yields 
\begin{equation}
\mathcal{A}=\bigcup_{r\in\mathbb{Z}}\left\{ \left(k_{x,}k_{y}\right)|r+\frac{1}{4}<\frac{k_{x}}{k_{\mathrm{max}}}<r+\frac{3}{4}\right\} ,
\end{equation}
and 
\begin{equation}
\sum_{\left(n,m\right)\in\mathcal{M}_{s}}\left|E_{n,m}^{\left(a\right)}\right|^{2}k_{\mathrm{min}}^{2}\approx F^{2}\int_{\mathcal{\mathcal{A}}}d^{2}\mathbf{k}\Phi\left(\mathbf{k}\right).\label{eq:A_integ_S8}
\end{equation}
Thus we obtain 
\begin{equation}
\left|\eta\right|\approx\sqrt{\frac{\int_{\mathcal{\mathcal{A}}}d^{2}\mathbf{k}\Phi\left(\mathbf{k}\right)}{\int_{\mathcal{\mathcal{S}}}d^{2}\mathbf{k}\Phi\left(\mathbf{k}\right)}}.\label{eq:absolute_eta_value2}
\end{equation}
We call the right hand side $\mu$ hereafter.

Figure 2 (b) is a conceptual picture relevant to the above argument,
which the reader may find useful. Figure 2 (c) shows $\mu$ as a function
of $\beta=\frac{\lambda}{2\sigma}$, where $\lambda=1.97\,\mathrm{pm}$
is the wavelength of $300\,\mathrm{keV}$ electrons.

\subsection{Multiple inelastic scattering in a single round of quantum measurement\label{subsec:Multiple-inelastic-scattering}}

Consider the effect of multiple inelastic scattering. Suppose that
they occurred $w$ times. We first note that, in the above entire
reasoning, the parameter $\alpha$ in Eq. (\ref{eq:generalized_initial_state})
could have been any complex number. We explicitly write $\alpha=\rho+i\kappa$,
and the state before inelastic scattering is 
\begin{equation}
|\overline{s}\rangle+\left(i\rho-\kappa\right)|\overline{a}\rangle.
\end{equation}
After inelastic scattering, we obtain 
\[
\left(|\overline{s}\rangle+i\alpha|\overline{a}\rangle\right)+\eta\left(|\overline{a}\rangle+i\alpha|\overline{s}\rangle\right)
\]
\begin{equation}
=\left(1+i\rho\eta-\kappa\eta\right)|\overline{s}\rangle+\left(i\rho-\kappa+\eta\right)|\overline{a}\rangle.
\end{equation}
Assuming that $\rho\eta$ and $\kappa\eta$ are small, this state
approximately equals 
\begin{equation}
|\overline{s}\rangle+\left(i\rho-\kappa+\eta\right)|\overline{a}\rangle.
\end{equation}
This shows that the effect of inelastic scattering additively accumulates.
In other words, after $n$ inelastic scattering, each with associated
$\eta$ parameter $\eta_{i},\,i=1,2,\cdots,n$, we obtain a Q1 state
of a form 
\begin{equation}
|\overline{s}\rangle+\left(i\rho+\sum_{i=1}^{n}\eta_{i}\right)|\overline{a}\rangle.\label{eq:Q1_after_multiple_inel_scatt}
\end{equation}
To develop a rough picture, assume that all $\eta_{i}$ have the same
absolute value $\mu>0$ derived in Eq. (\ref{eq:absolute_eta_value2}),
but their signs are random. Then $\sum_{i=1}^{w}\eta_{i}$ represents
random walk on the real line, which results in 
\begin{equation}
\left|\sum_{i=1}^{w}\eta_{i}\right|\approx\sqrt{w}\mu.\label{eq:Q1_after_multiple_inel_scatt2}
\end{equation}

\subsection{Estimating the final outcome of imaging\label{subsec:Estimating-the-final}}

Finally, we consider how the accumulated error $\sqrt{w}\mu$ affects
our measurement. Consider a general qubit state 
\begin{equation}
|\overline{\psi}\rangle=e^{-i\frac{\varphi}{2}}\cos\frac{\theta}{2}|\overline{0}\rangle+e^{i\frac{\varphi}{2}}\sin\frac{\theta}{2}|\overline{1}\rangle,
\end{equation}
where $\theta,\,\varphi$ are latitude and longitude on the Bloch
sphere, respectively. Comparing with Eq. (\ref{eq:Q1_after_multiple_inel_scatt},
\ref{eq:Q1_after_multiple_inel_scatt2}), we find that $\theta,\,\varphi$
correspond to $\rho,\,\sqrt{w}\mu$ as 
\begin{equation}
\varphi=-2\rho,
\end{equation}
\begin{equation}
\theta=\frac{\pi}{2}-2\sqrt{w}\mu=\frac{\pi}{2}-\Theta,
\end{equation}
where we defined $\Theta$ to indicate the angular deviation from
the ideal great circle on the Bloch sphere, which passes the states
$|\overline{s}\rangle,\,|\overline{a}\rangle,\,|\overline{\uparrow}\rangle,\,|\overline{\downarrow}\rangle$.
Hence $|\overline{\psi}\rangle$ expressed in terms of $\theta,\,\varphi$
is 
\begin{equation}
|\overline{\psi}\rangle=e^{i\rho}\cos\left(\frac{\pi}{4}-\frac{\Theta}{2}\right)|\overline{0}\rangle+e^{-i\rho}\sin\left(\frac{\pi}{4}-\frac{\Theta}{2}\right)|\overline{1}\rangle.
\end{equation}
Using the basis state $|\overline{\uparrow}\rangle,\,|\overline{\downarrow}\rangle$,
we obtain (multiplying an overall phase factor $e^{i\frac{\varphi}{2}}$)
\begin{equation}
|\overline{\psi}\rangle=C_{\uparrow}|\overline{\uparrow}\rangle+C_{\downarrow}|\overline{\downarrow}\rangle,
\end{equation}
where 
\begin{equation}
C_{\uparrow}=\frac{\cos\frac{\theta}{2}-ie^{i\varphi}\sin\frac{\theta}{2}}{\sqrt{2}},
\end{equation}
\begin{equation}
C_{\downarrow}=\frac{\cos\frac{\theta}{2}+ie^{i\varphi}\sin\frac{\theta}{2}}{\sqrt{2}}
\end{equation}
and hence 
\[
p_{\uparrow}=\left|C_{\uparrow}\right|^{2}=\frac{1+\sin\theta\sin\varphi}{2}
\]
\begin{equation}
=\frac{1-\cos\Theta\sin\left(2\rho\right)}{2}\approx\frac{1}{2}-\rho\cos\Theta,
\end{equation}

\begin{equation}
p_{\downarrow}=\left|C_{\downarrow}\right|^{2}\approx\frac{1}{2}+\rho\cos\Theta.
\end{equation}
Hence the signal we want to detect, $\rho$, weakens by the factor
$\cos\Theta$.

\begin{figure}[t]
\includegraphics[scale=0.35]{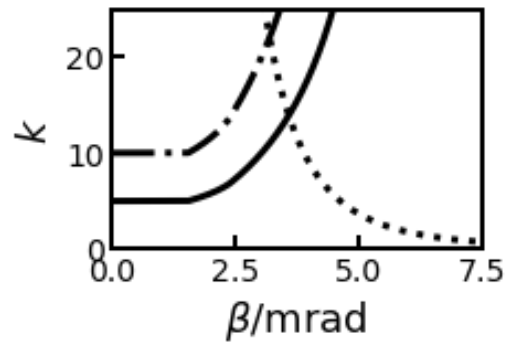}

\caption{The optimal repetition number $k$, enabled by inelastic scattering
neutralization, is plotted against $\beta=\lambda/2\sigma$, where
$\sigma$ is resolution of interest. For example, $\beta=2\,\mathrm{mrad}$
corresponds to $\sigma=0.5\,\mathrm{nm}$. The phase resolution is
improved by $\approx0.65\sqrt{k}$. The dash-dotted curve and the
solid curve correspond to $k_{\mathrm{opt}}$ defined in the main
text for $\Lambda/t=10$ and $5$, which roughly correspond to the
specimen thickness of $t\approx30\,\mathrm{nm}$ and $60\,\mathrm{nm}$,
respectively. On the other hand, radiation damage governs the allowed
number of electrons $N_{\mathrm{sq}}$ (the dotted curve) and the
actual repetition number $k$ should equal $\min\left\{ k_{\mathrm{opt}},\,N_{\mathrm{sq}}\right\} $.}
\end{figure}

Next, we consider improvement in terms of signal-to-noise ratio (SNR).
A single round of quantum measurement comprises transmission of $k$
electrons through the specimen. The number of inelastic scattering
events in a single round of quantum measurement is, on average 
\begin{equation}
w=\frac{kt}{\Lambda},
\end{equation}
where $k$ is the repetition number, $t$ is the specimen thickness,
and $\Lambda$ is the inelastic mean free path. Let half the phase
difference between specimen areas $0$ and $1$, where the electron
states $|\overline{0}\rangle,\,|\overline{1}\rangle$ are respectively
focused, that we want to measure, be $\overline{\theta}$. The accumulated
phase after a single round of quantum measurement, weakened by inelastic
scattering events, is 
\begin{equation}
k\overline{\theta}\cos\Theta=k\overline{\theta}\cos\left(2\sqrt{w}\mu\right)=k\overline{\theta}\cos\left(2\sqrt{\frac{kt}{\Lambda}}\mu\right).\label{eq:signal_amplitude}
\end{equation}
We use $N$ electrons and try this $\frac{N}{k}$ times. In the language
of binomial distribution $B\left(n,p\right)$, where $n=\frac{N}{k}$
and $p=\frac{1}{2}+k\overline{\theta}\cos\Theta$ in the present case,
the mean is $np=\frac{N}{2k}+N\overline{\theta}\cos\Theta$ and the
variance is $np\left(1-p\right)\approx\frac{N}{4k}$. We divide the
signal 
\begin{equation}
N\overline{\theta}\cos\Theta\label{eq:signal_S8}
\end{equation}
by noise, which is square root of the variance 
\begin{equation}
\frac{1}{2}\sqrt{\frac{N}{k}}.\label{eq:noise_S8}
\end{equation}
We want to find the optimal repetition number $k_{2}$, which maximizes
SNR. To find $k_{2}$, we may consider a quantity that is square of
the ratio of Eq. (\ref{eq:signal_S8}) to Eq. (\ref{eq:noise_S8}):
\begin{equation}
4Nk\overline{\theta}^{2}\cos^{2}\Theta\propto k\cos^{2}\left(2\sqrt{\frac{kt}{\Lambda}}\mu\right)=F\left(k\right).
\end{equation}
Then, we should get $\frac{dF\left(k\right)}{dk}=0$ at $k=k_{2}$.
We find a condition 
\[
\frac{dF\left(k\right)}{dk}=\cos^{2}\xi-\xi\cos\xi\sin\xi=0,
\]
or equivalently 
\begin{equation}
\frac{1}{\xi}=\tan\xi,\label{eq:tangent_relation}
\end{equation}
in terms of 
\begin{equation}
\xi=2\sqrt{\frac{kt}{\Lambda}}\mu.
\end{equation}
Numerical solution to this equation turns out to be $\xi=0.86$ and
we further obtain $\xi^{2}=0.74$. Thus we get 
\begin{equation}
k_{2}=\frac{\xi^{2}}{4\mu^{2}}\frac{\Lambda}{t}\approx\frac{0.74}{4\mu^{2}}\frac{\Lambda}{t}\label{eq:optimal_k}
\end{equation}
and improvement of S/N is 
\begin{equation}
\sqrt{k_{2}}\cos\left(2\sqrt{\frac{k_{2}t}{\Lambda}}\mu\right)=\sqrt{k_{2}}\cos\xi.\label{eq:SN_improvement}
\end{equation}
When $\mu^{2}$ is large at low resolution, we may as well not perform
inelastic scattering neutralization. In the absence of inelastic scattering
neutralization, the optimal $k$ is $k_{1}=\Lambda/t$ and improvement
of S/N ratio is $\sqrt{k_{1}/e}=\sqrt{\Lambda/et}$ as shown in Eq.
(\ref{eq:garden_variety_QEM_improvement}). These results show advantage
of inelastic scattering neutraliztion because we can use a larger
$k_{2}$ than $k_{1}$ when $\mu^{2}$ is small, as shown in Eq. (\ref{eq:optimal_k}).
Otherwise we should employ $k_{1}$. In this case, S/N ratios given
by Eqs. (\ref{eq:garden_variety_QEM_improvement}) and (\ref{eq:SN_improvement})
turn out to be similar. Indeed, numerically, $\cos\xi\approx0.65$
is close to $\frac{1}{\sqrt{e}}\approx0.61$, indicating that Eq.
(\ref{eq:SN_improvement}) is not too bad with a moderately large
$\mu^{2}$, although our calculations implicitly assumed a small $\mu^{2}$.
See Sec. \ref{sec:Expected-improvement} for information about the
repetition number $k$ that the experimenter should choose at a given
SF.

Since it is better \emph{not} to use ISN when $\mu^{2}$ is too large,
we define $k_{\mathrm{opt}}=\max\left\{ k_{2},k_{1}\right\} $. On
the other hand, $k$ cannot exceed the dose limit $N_{\mathrm{sq}}$
of Eq. (\ref{eq:quartic_dose}). Figure 3 shows how $k_{\mathrm{opt}}$
and $N_{\mathrm{sq}}$ depend on $\beta=\lambda/2\sigma$.

\section{Expected improvement: A simulation study\label{sec:Expected-improvement}}

To visually assess the improvement afforded by ISN, we simulate imaging
of the Marburg virus VP35 domain molecule \cite{5TOI}. Figure 4 shows
the result. The noiseless map of phase shift shown in Fig. 4 (a) is
produced by multislice simulation as described in Appendix D. Figures
4 (b)-(f) are addition of the phase map and noise. Figure 4 (b) shows
the case without quantum advantage. (However, SF-dependent electron
dose control mentioned in Sec. \ref{subsec:High-level-ideas} is employed
and hence this is \emph{not} conventional TEM imaging.) Figures 4
(c), (d) shows simulated images with and without ISN for a thin ($\approx30\,\mathrm{nm}$)
specimen, whereas Figs. 4 (e), (f) shows corresponding images for
a relatively thick ($\approx60\,\mathrm{nm}$) specimen. For simplicity,
we assume that energy loss $E$ is always $20\,\mathrm{eV}$ \cite{Leapman_Sun}.
All computations are performed on $240\times240$ pixels image data,
with each square pixel having the side length $l=0.05\,\mathrm{nm}$.
We label each pixel with a pair of integers $\left(n,\,m\right)$,
each of which ranges from $-120$ to $119$.

\begin{figure}[t]
\includegraphics[scale=0.22]{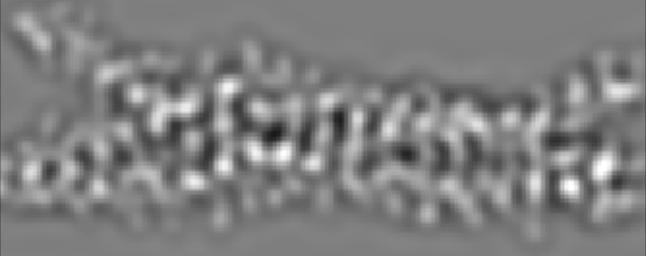} (a) \includegraphics[scale=0.22]{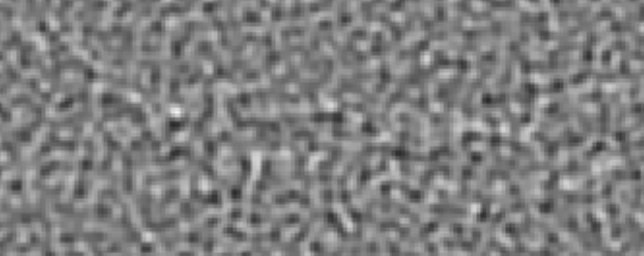}
(b) \includegraphics[scale=0.22]{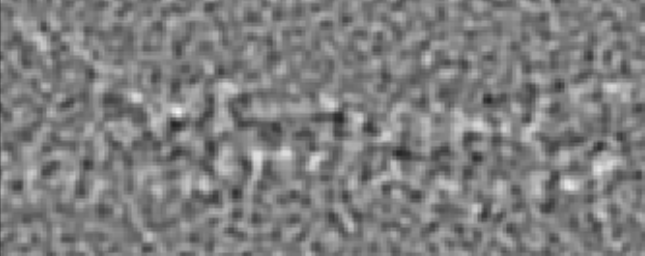} (c) \includegraphics[scale=0.22]{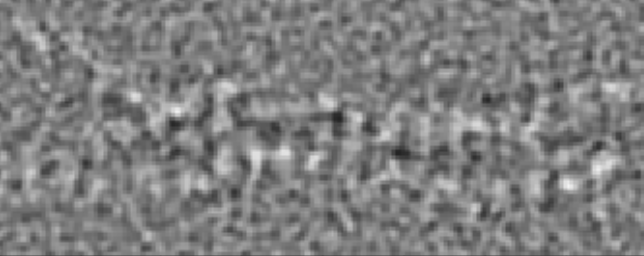}
(d)

\includegraphics[scale=0.22]{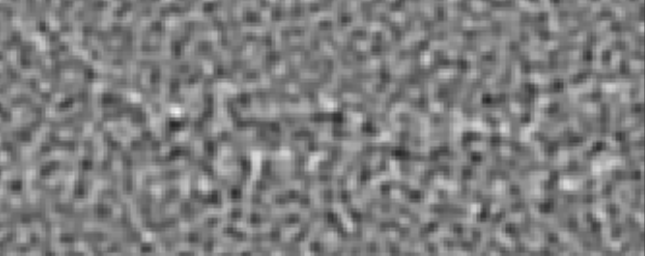} (e) \includegraphics[scale=0.22]{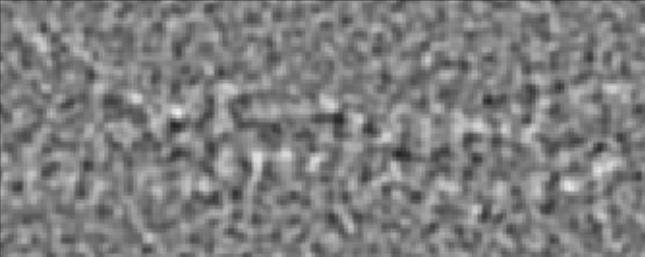}
(f)

\caption{Simulated images of the Marburg virus VP35 domain molecule \cite{5TOI}.
(a) A band-pass filtered phase map. (b) Imaging without quantum enhancement,
i.e. the repetition number $k=1$. (c) QEM imaging without inelastic
scattering neutralization (ISN), wherein $k_{1}=\Lambda/t=10$, roughly
corresponding to a specimen thickness $t\approx30\,\mathrm{nm}$.
(d) QEM imaging with ISN. The repetition number is $\tilde{k}_{2}$,
where $\Lambda/t=10$. (e) QEM imaging without ISN, wherein $\Lambda/t=5$
(i.e. specimen thickness $t\approx60\,\mathrm{nm}$). (f) QEM imaging
with ISN, wherein $\Lambda/t=5$. The horizontal length of all images
is $10\,\mathrm{nm}$.}
\end{figure}

In what follows, we describe the procedure to generate noise in each
case. Typically, although not always, many rounds of quantum measurements,
each involving $k$ electron passing events, are performed for each
pixel. Hence we expect the noise to be approximately gaussian, which
approximates the binomial distribution. In all three imaging methods
--- i.e. TEM with no quantum enhancement but with SF-dependent dose
control, ``conventional QEM'', and QEM with ISN --- the amount
of noise depends on 
\begin{equation}
\beta=\frac{q}{k_{z}}=\frac{\pi}{\sigma k_{z}}=\frac{\lambda}{2\sigma},\label{eq:beta_q_relation_S10}
\end{equation}
where $k_{z}$ is the wave number along the optical axis, and $\lambda=2\pi/k_{z}\approx1.97\,\mathrm{pm}$
is the wavelength of $300\,\mathrm{keV}$ electrons. Equation (\ref{eq:beta_q_relation_S10})
is valid insofar as the vector $\mathbf{q}$ may be regarded as being
perpendicular to the optical axis.

Specific steps are the following. First, we generate real-valued,
independent gaussian noise, with zero mean and unit variance, in each
pixel on the image plane. Second, we perform fast Fourier transform
(FFT) to obtain the noise in the diffraction plane, which results
in a complex-valued map. The pixel $\left(0,119\right)$ in the map,
for example, corresponds to a scattering angle 
\begin{equation}
\beta\approx\sin\beta=\frac{\lambda}{2l}=\frac{1.97\,\mathrm{pm}}{2\times0.05\,\mathrm{nm}}\times\frac{119}{120}=19.5\,\mathrm{mrad}.
\end{equation}
Third, to the map on the diffraction plane we multiply a function
that describes the $q$-dependent amplitude of noise. For the ``classical''
case of Fig. 4 (b), we multiply the standard deviation of shot noise
\[
\frac{1}{\sqrt{N_{\mathrm{sq}}}}=\frac{1}{\sqrt{F_{\mathrm{sq}}A}}=\sqrt{\frac{Rq^{4}}{32\pi^{5}\zeta}}=\sqrt{\frac{R}{32\pi^{5}\zeta}}\frac{q^{2}}{k_{z}^{2}}k_{z}^{2}
\]
\begin{equation}
=\frac{10^{-6}}{\lambda^{2}}\sqrt{\frac{R}{2\pi\zeta}}\left(\frac{\beta}{\mathrm{mrad}}\right)^{2}=\frac{\left(\beta/\mathrm{mrad}\right)^{2}}{186},\label{eq:classical_noise_S10}
\end{equation}
where $\zeta=0.255$ as described in Sec. \ref{subsec:High-level-ideas}.
Uncertainties associated with parameters such as $R,\,\zeta$ do not
warrant the precision appearing in the numerical value $186$, but
we use this value in the simulation anyway. Equation (\ref{eq:classical_noise_S10})
overestimates noise when $N_{\mathrm{sq}}<1$, where we do not perform
measurement and hence there is no noise (and no signal). However,
computed images in Fig. 4 is bandpass-filtered anyway and hence this
particular artefact does not matter. 

The images shown in Figs. 4 (c)-(f) have smaller noise than Eq. (\ref{eq:classical_noise_S10})
indicates, because of quantum advantage. To discuss degrees of noise
reduction, using $k_{1},k_{\mathrm{opt}}$ and $\xi$ defined in Sec.
\ref{subsec:Estimating-the-final}, we define 
\begin{equation}
\tilde{k}_{1}=\max\left\{ \min\left\{ k_{1},N_{\mathrm{sq}}\right\} ,e\right\} ,\label{eq:k1_tilda}
\end{equation}
and 
\begin{equation}
\tilde{k}_{2}=\max\left\{ \min\left\{ k_{\mathrm{opt}},N_{\mathrm{sq}}\right\} ,\cos^{-2}\xi\right\} .\label{eq:k2_tilda}
\end{equation}
We obtain $\cos^{-2}\xi=2.35$ from Eq. (\ref{eq:tangent_relation}).
Hence Eq. (\ref{eq:k2_tilda}) can more specifically be written as,
noting $\sqrt{0.74/4}=0.43$,
\begin{equation}
\tilde{k}_{2}=\max\left\{ \min\left\{ k_{1}\max\left\{ \left(\frac{0.43}{\mu}\right)^{2},1\right\} ,N_{\mathrm{sq}}\right\} ,2.35\right\} .\label{eq:k2_tilda2}
\end{equation}
Remark: Figure 2 (c), along with Fig. 3, clearly show that the resolution,
at which $N_{\mathrm{sq}}=2.35$, is much higher than another resolution,
where $\mu=0.43$ holds.

For the QEM cases, we multiply factors of noise reduction compared
to the classical case of Eq. (\ref{eq:classical_noise_S10}). They
are, first, 
\begin{equation}
\sqrt{\frac{e}{\tilde{k}_{1}}}.
\end{equation}
for QEM without ISN and second, 
\begin{equation}
\frac{1}{\sqrt{\tilde{k}_{2}}\cos\xi}
\end{equation}
for QEM with ISN.

Here we elaborate on Eq. (\ref{eq:k1_tilda}) somewhat. Since the
noise reduction factor compared to classical imaging is $\sqrt{\frac{e}{\tilde{k}_{1}}}$,
there is no point in employing $\tilde{k}_{1}$ less than $e$. One
would simply perform classical measurement in this case. The outermost
``max'' function in Eq. (\ref{eq:k1_tilda}) ensures that we get
at least the classical performance. Otherwise we would use $k_{1}$
as the repetition number, \emph{unless} $k_{1}$ exceeds the dose
limit $N_{\mathrm{sq}}$. In the latter case, we take $N_{\mathrm{sq}}$
as the repetition number of the quantum measurement.

The idea behind Eq. (\ref{eq:k2_tilda2}) for $\tilde{k}_{2}$ is
similar. Once again, the outermost ``max'' function ensures the
classical performance in the worst case. The ``min'' function in
it ensures that we do not exceed the dose limit $N_{\mathrm{sq}}$.
The innermost ``max'' function makes sure that ISN is employed only
when it is advantageous, compared to ``conventional'' QEM, to do
so.

Fourth and finally, we apply inverse FFT to obtain spatial-frequency-weighted
noise patterns. The result should mathematically be real, but the
real part should be taken in actual numerical computation. The resultant
noise patterns are simply added to the phase map, i.e. $\theta\left(x,y\right)$
of the exit wave $1+i\theta\left(x,y\right)$ (shown in Fig. 4 (a))
to obtain Figs. 4 (b)-(f).

A bandpass filter is applied to images because the improvement by
ISN is mainly at high resolution and visually rather subtle, requiring
removal of large low resolution components. Thus all six images are
filtered by multiplying a function $e^{-\beta^{2}/2\beta_{H}^{2}}\left(1-e^{-\beta^{2}/2\beta_{L}^{2}}\right)$
in the $\beta_{x}\beta_{y}$-space, where $\beta^{2}=\beta_{x}^{2}+\beta_{y}^{2}$,
$\beta_{L}=2\,\mathrm{mrad}$ and $\beta_{H}=3.5\,\mathrm{mrad}$.
The contrast of all images in Fig. 4 (a)-(f) are adjusted in the following
way. Given a numerical array representing an image, the mean $\bar{\mu}$
and the standard deviation $\bar{\sigma}$ are computed. The highest
and the lowest brightness in each presented image are then made to
correspond to the values $\bar{\mu}+5\bar{\sigma}$ and $\bar{\mu}-5\bar{\sigma}$,
respectively. Finally, the images are cropped to the size of approximately
$80\times200$ pixels for presentation.

\section{Conclusion}

In summary, it is in principle possible to neutralize, to the extent
we discussed, inelastic scattering especially at high resolution.
We conjecture that this is essentially the fundamental limit of electron
microscopy of beam sensitive specimens, when performing standard imaging.
On the other hand, non-standard imaging, such as image verification
\cite{OLF} among other possibilities, is worth further study.

\section*{ACKNOWLEDGMENT}

The author thanks Professors R. F. Egerton and M. Malac for useful
discussions on inelastic scattering. This research was supported in
part by the JSPS ``Kakenhi'' Grant (Grant No. 19K05285).

\section*{APPENDIX A: ENTANGLEMENT ENHANCED ELECTRON MICROSCOPY: A BRIEF INTRODUCTION}

In this section, we briefly review entanglement-enhanced electron
microscopy \cite{eeem_cpb,eeem_flux,eeem_error} for readers who are
unfamiliar with the scheme. This review purposely avoids the implementation
aspect of the scheme, such as the use of superconducting quantum devices.
Instead, we focus on principles and therefore we take it for granted
that all theoretically possible operations, such as unitary transformations,
addition of an initialized ancilla qubit, and measurements on arbitrary
subsystems are possible. We ignore the spin of the electron, regarding
it to be decoupled from the degrees of freedom of interest. We intend
to make Appendix A as self-contained as possible. As a result, there
are few redundancies with the main text.

Let the electron states $|0\rangle,\,|1\rangle$ be ones that are
localized on non-overlapping regions $0$ and $1$ on a biological
specimen when the electron passes the specimen. Recall that the specimen
may be regarded as a weak phase object in cryoEM. We want to measure
the difference in phase shifts between the regions $0$ and $1$.
Keep in mind that there are many electron states other than $|0\rangle,\,|1\rangle$.
Define symmetric and asymmetric states as 
\begin{equation}
|s\rangle=\frac{|0\rangle+|1\rangle}{\sqrt{2}},\;|a\rangle=\frac{|0\rangle-|1\rangle}{\sqrt{2}}.\label{eq:s_a_state_def_S7}
\end{equation}
Let the initial electron state be $|0\rangle$. Consider a separate
$2$-state system (call it qubit Q1) with basis states $|\overline{0}\rangle,\,|\overline{1}\rangle$,
where the bar indicates that the state belongs to Q1. Another set
of basis states $|\overline{s}\rangle,\,|\overline{a}\rangle$ for
Q1 is defined in terms of $|\overline{0}\rangle,\,|\overline{1}\rangle$
exactly as in Eq. (\ref{eq:s_a_state_def_S7}). Suppose that Q1 is
in the state 
\begin{equation}
|\overline{\alpha}\rangle=\frac{e^{i\alpha}|\overline{0}\rangle+e^{-i\alpha}|\overline{1}\rangle}{\sqrt{2}},\label{eq:qubit_initial_S7}
\end{equation}
for some $\alpha$, whose significance will be apparent shortly. We
occasionally write a state of the combined system of an electron in
a state $u$, and the qubit in a state $\overline{v}$, as 
\begin{equation}
|u\overline{v}\rangle=|u\rangle\otimes|\overline{v}\rangle.
\end{equation}
Hence we write the current state as $|0\overline{\alpha}\rangle$.
Let $U$ be a unitary operation, which flips the electron state as
$|0\rangle\Rightarrow|1\rangle,\,|1\rangle\Rightarrow|0\rangle$ if
and only if the qubit is in the state $|\overline{1}\rangle$. (This
should be realizable with a superconducting qubit \cite{eeem_cpb,eeem_flux,eeem_error}.)
In other words, 
\begin{equation}
\alpha|0\rangle+\beta|1\rangle\Rightarrow\beta|0\rangle+\alpha|1\rangle
\end{equation}
if and only if the qubit is in the state $|\overline{1}\rangle$.
In short, $U$ is the quantum controlled-NOT gate in the language
of quantum information science. We apply $U$ to the electron-qubit
system in the initial state $|0\overline{\alpha}\rangle$, which results
in 
\begin{equation}
\frac{e^{i\alpha}|0\overline{0}\rangle+e^{-i\alpha}|1\overline{1}\rangle}{\sqrt{2}}.\label{eq:combined_initial_S7}
\end{equation}
Upon passing the biological specimen, the two electron states $|0\rangle,\,|1\rangle$
acquire different phase shifts as 
\begin{equation}
|0\rangle\Rightarrow e^{i\theta}|0\rangle,\;|1\rangle\Rightarrow e^{-i\theta}|1\rangle,\label{eq:phase_shift_S7}
\end{equation}
where $\theta$ is half the relative phase shift between the specimen
regions $0$ and $1$, which correspond respectively to the localized
electron states $|0\rangle,\,|1\rangle$. Our objective is to determine
$\theta$ as precisely as possible. After the electron transmits through
the specimen, the state of the entire system is \begin{widetext}
\begin{equation}
\frac{e^{i\left(\alpha+\theta\right)}|0\overline{0}\rangle+e^{-i\left(\alpha+\theta\right)}|1\overline{1}\rangle}{\sqrt{2}}=\frac{|s\rangle}{\sqrt{2}}\otimes\frac{e^{i\left(\alpha+\theta\right)}|\overline{0}\rangle+e^{-i\left(\alpha+\theta\right)}|\overline{1}\rangle}{\sqrt{2}}+\frac{|a\rangle}{\sqrt{2}}\otimes\frac{e^{i\left(\alpha+\theta\right)}|\overline{0}\rangle-e^{-i\left(\alpha+\theta\right)}|\overline{1}\rangle}{\sqrt{2}}.
\end{equation}
\end{widetext}Next, we measure the electron state in the basis $\left\{ |s\rangle,\,|a\rangle\right\} $.
If the measurement outcome indicates $|s\rangle$, then the qubit
is left in the state

\begin{equation}
\frac{e^{i\left(\alpha+\theta\right)}|\overline{0}\rangle+e^{-i\left(\alpha+\theta\right)}|\overline{1}\rangle}{\sqrt{2}}.\label{eq:qubit_final_S7}
\end{equation}
Likewise, when the outcome indicates $|a\rangle$, then the qubit
state is 
\begin{equation}
\frac{e^{i\left(\alpha+\theta\right)}|\overline{0}\rangle-e^{-i\left(\alpha+\theta\right)}|\overline{1}\rangle}{\sqrt{2}},
\end{equation}
which can readily be brought to Eq. (\ref{eq:qubit_final_S7}) by
the Pauli-Z-gate operation $|\overline{0}\rangle\Rightarrow|\overline{0}\rangle,\;|\overline{1}\rangle\Rightarrow-|\overline{1}\rangle$.
Thus, the overall effect of passing an electron through the specimen
is to have the qubit state evolution from Eq. (\ref{eq:qubit_initial_S7})
to Eq. (\ref{eq:qubit_final_S7}). This means that we can start with
$\alpha=0$ and repeat the process $k$ times, which means that $k$
electrons pass the specimen, to obtain the state 
\begin{equation}
\frac{e^{ik\theta}|\overline{0}\rangle+e^{-ik\theta}|\overline{1}\rangle}{\sqrt{2}}.\label{eq:accumulated_phase_S7}
\end{equation}
Consequently, the small phase shift \emph{$\theta$} accumulate on
Q1 $k$ times. Next, we measure this qubit state with respect to the
basis states 
\begin{equation}
|\overline{\uparrow}\rangle=\frac{|\overline{0}\rangle+i|\overline{1}\rangle}{\sqrt{2}},\,|\overline{\downarrow}\rangle=\frac{|\overline{0}\rangle-i|\overline{1}\rangle}{\sqrt{2}}.
\end{equation}
Hence Eq. (\ref{eq:accumulated_phase_S7}) equals \begin{widetext}
\[
\frac{1}{2}\left\{ e^{ik\theta}\left(|\overline{\uparrow}\rangle+|\overline{\downarrow}\rangle\right)-ie^{-ik\theta}\left(|\overline{\uparrow}\rangle-|\overline{\downarrow}\rangle\right)\right\} =\frac{e^{ik\theta}-ie^{-ik\theta}}{2}|\overline{\uparrow}\rangle+\frac{e^{ik\theta}+ie^{-ik\theta}}{2}|\overline{\downarrow}\rangle
\]
\begin{equation}
=e^{-i\frac{\pi}{4}}\frac{e^{i\left(\frac{\pi}{4}+k\theta\right)}+e^{-i\left(\frac{\pi}{4}+k\theta\right)}}{2}|\overline{\uparrow}\rangle+e^{i\frac{\pi}{4}}\frac{e^{i\left(\frac{\pi}{4}-k\theta\right)}+e^{-i\left(\frac{\pi}{4}-k\theta\right)}}{2}|\overline{\downarrow}\rangle.
\end{equation}
\end{widetext}Hence probabilities for the two outcomes $\uparrow,\,\downarrow$
are 
\begin{equation}
p_{\uparrow}=\cos^{2}\left(\frac{\pi}{4}+k\theta\right),\,p_{\downarrow}=\cos^{2}\left(\frac{\pi}{4}-k\theta\right).
\end{equation}
Hence we obtain 
\begin{equation}
p_{\uparrow}=\frac{1-\sin2k\theta}{2},\,p_{\downarrow}=\frac{1+\sin2k\theta}{2}.\label{eq:probabilities_S7}
\end{equation}
(Further consideration shows that the Pauli-Z gate operations mentioned
above does not have to be performed at all, as long as we count the
number of measurement outcomes corresponding to $|a\rangle$.) This
form of probability offers advantage over the conventional method
that use $k$ electrons separately, i.e. a set of $k$ measurements,
each with 
\begin{equation}
\hat{p}_{\uparrow}=\frac{1-\sin2\theta}{2},\,\hat{p}_{\downarrow}=\frac{1+\sin2\theta}{2}.\label{eq:probabilities2_S7}
\end{equation}
for the following reasons. Assume $k\theta\ll1$ for simplicity and
consider the quantum-enhanced case represented with Eq. (\ref{eq:probabilities_S7})
first. Let a random variable $X$ be such that its value is $1$ when
the measurement outcome is $\uparrow$ and otherwise $0$. The expectation
value of $X$ is $\left\langle X\right\rangle =p_{\uparrow}=\left(1-2k\theta\right)/2$
while the standard deviation is $\sigma\left(X\right)=\sqrt{p_{\uparrow}p_{\downarrow}}\approx1/2$.
Consider another random variable $Y=\frac{1-2X}{2k}$, which is designed
to have the property $\left\langle Y\right\rangle =\theta$. Its standard
deviation is $\sigma\left(Y\right)\approx\frac{1}{2k}$ because $\sigma\left(Y\right)=\sqrt{\mathrm{Var}\left(Y\right)}$
and 
\[
\mathrm{Var}\left(Y\right)=\mathrm{Var}\left(\frac{1-2X}{2k}\right)
\]
\begin{equation}
=\frac{1}{k^{2}}\mathrm{Var}\left(X\right)=\frac{1}{k^{2}}\sigma^{2}\left(X\right)=\frac{1}{4k^{2}}.
\end{equation}
On the other hand, consider the ``classical'' case, where a measurement
represented by Eq. (\ref{eq:probabilities2_S7}) is repeated $k$
times. A relevant random variable here is $Z\sim B\left(k,\hat{p}_{\uparrow}\right)$,
where $B\left(k,\hat{p}_{\uparrow}\right)$ is the binomial distribution,
with $\left\langle Z\right\rangle =k\hat{p}_{\uparrow}=k\left(1-2\theta\right)/2$
and $\sigma\left(Z\right)=\sqrt{k\hat{p}_{\uparrow}\hat{p}_{\downarrow}}\approx\sqrt{k}/2$.
Take another random variable $W=\frac{1}{2}-\frac{Z}{k}$ designed
for the property $\left\langle W\right\rangle =\theta$. One finds
$\sigma\left(W\right)\approx\frac{1}{2\sqrt{k}}$, which is worse
than the quantum-enhanced case.

It is instructive to view the exact same measurement process from
the perspective of another basis $\left\{ |s\rangle,\,|a\rangle\right\} $
and $\left\{ |\overline{s}\rangle,\,|\overline{a}\rangle\right\} $.
The Q1 state $|\overline{\sigma}\rangle$ in Eq. (\ref{eq:qubit_initial_S7})
is expressed as 
\begin{equation}
|\overline{\sigma}\rangle=\cos\alpha|\overline{s}\rangle+i\sin\alpha|\overline{a}\rangle.
\end{equation}
The initial electron state is 
\begin{equation}
|0\rangle=\frac{|s\rangle+|a\rangle}{\sqrt{2}}
\end{equation}
As is well-known and readily verifiable, roles of the control qubit
and the target qubit of the controlled-NOT gate are swapped upon the
change of the basis states. The controlled-NOT $U$ flips the \emph{qubit}
state as $|\overline{s}\rangle\Rightarrow|\overline{a}\rangle,\,|\overline{a}\rangle\Rightarrow|\overline{s}\rangle$
if and only if the \emph{electron} is in the state $|a\rangle$. Applying
$U$ to the combined initial state $|0\alpha\rangle$, we obtain 
\begin{equation}
\cos\alpha\left(\frac{|s\overline{s}\rangle+|a\overline{a}\rangle}{\sqrt{2}}\right)+i\sin\alpha\left(\frac{|s\overline{a}\rangle+|a\overline{s}\rangle}{\sqrt{2}}\right).\label{eq:combined_initial2_S7}
\end{equation}
Equation (\ref{eq:phase_shift_S7}) now states 
\begin{equation}
|s\rangle\Rightarrow\cos\theta|s\rangle+i\sin\theta|a\rangle,\;|a\rangle\Rightarrow\cos\theta|a\rangle+i\sin\theta|s\rangle.\label{eq:basic_scattering_S7}
\end{equation}
Hence, after the electron passes the specimen, the entire state is
\begin{widetext} 
\[
\cos\theta\left\{ \cos\alpha\left(\frac{|s\overline{s}\rangle+|a\overline{a}\rangle}{\sqrt{2}}\right)+i\sin\alpha\left(\frac{|s\overline{a}\rangle+|a\overline{s}\rangle}{\sqrt{2}}\right)\right\} +i\sin\theta\left\{ \cos\alpha\left(\frac{|s\overline{a}\rangle+|a\overline{s}\rangle}{\sqrt{2}}\right)+i\sin\alpha\left(\frac{|s\overline{s}\rangle+|a\overline{a}\rangle}{\sqrt{2}}\right)\right\} 
\]
\begin{equation}
=\cos\left(\alpha+\theta\right)\left(\frac{|s\overline{s}\rangle+|a\overline{a}\rangle}{\sqrt{2}}\right)+i\sin\left(\alpha+\theta\right)\left(\frac{|s\overline{a}\rangle+|a\overline{s}\rangle}{\sqrt{2}}\right).
\end{equation}
\end{widetext}We measure the electron state in the basis $\left\{ |s\rangle,\,|a\rangle\right\} $.
If the measurement outcome is $|s\rangle$, then the qubit is left
in the state 
\begin{equation}
\cos\left(\alpha+\theta\right)|\overline{s}\rangle+i\sin\left(\alpha+\theta\right)|\overline{a}\rangle.\label{eq:qubit_final2_S7}
\end{equation}
Likewise, if the outcome is $|a\rangle$, then we obtain 
\begin{equation}
\cos\left(\alpha+\theta\right)|\overline{a}\rangle+i\sin\left(\alpha+\theta\right)|\overline{s}\rangle.
\end{equation}
This state can readily be brought to the form Eq. (\ref{eq:qubit_final2_S7})
by the operation $|\overline{s}\rangle\Longleftrightarrow|\overline{a}\rangle$.
The measurement of Q1 after passing $k$ electrons proceeds in the
same way described in the above. In summary, we see that, in this
alternative basis, the electron wave gets scattered from the state
$|s\rangle$ to $|a\rangle$ or \emph{vice versa} with a small amplitude
$\approx i\theta$. This amplitude accumulates in the state $|\overline{a}\rangle$
of Q1.

Finally, we consider processes that lead to failure of the measurement.
The first such process is inelastic electron scattering. In the simplest
case, the probe electron excites a localized degree of freedom of
the specimen. This leads to, for example, ejection of a K-shell electron
in the specimen. The probe electron position is effectively ``measured''
in this process because the excitation is localized within the region
$0$ \emph{or} $1$. Hence, the electron is projected onto a state
that has overlap with only one of the states $|0\rangle$ and $|1\rangle$.
As a result, the state Eq. (\ref{eq:combined_initial_S7}) gets disentangled
and the qubit state is projected onto $|\overline{0}\rangle$ or $|\overline{1}\rangle$.
Thus, we completely lose the information encoded in the parameter
$\alpha$. We waste all the dose budget corresponding to $\kappa\le k$
electrons if such inelastic scattering happens after using $\kappa$
electrons in the round of measurement. Fortunately, K-shell ejection
processes have small scattering cross sections in cryoEM \cite{eeem_error}.
A somewhat more delocalized plasmon excitations are much more frequent.
(The typical energy loss due to plasmons is $\Delta E\approx20\,\mathrm{eV}$
\cite{Leapman_Sun}.) The problem is less severe at \emph{higher}
resolution, where regions $0$ and $1$ are close, because both the
states $|0\rangle$ and $|1\rangle$ may be within the delocalization
length. In the far field, the degree of localization manifests itself
as the angular spread of the inelastically scattered wave. For example,
if excitation of an atom caused localization of the electron wave
to an atomic dimension $\delta x$, then the spread of the scattered
wave $\sim\lambda/\delta x$ would be much larger than what is observed.
Our hope is to keep the absolute amplitudes pertaining to $|\overline{0}\rangle$
and $|\overline{1}\rangle$ balanced after an inelastic scattering
event. In the present work, we wish to determine whether the detected
electron originates from the state $|s\rangle$ or $|a\rangle$, in
spite of the angular spread caused by inelastic scattering. See the
main text for our strategy.

The second process that may lead to failure of measurement is \emph{elastic}
scattering. This process involves electron states outside the Hilbert
space spanned by $|0\rangle$ and $|1\rangle$. Call them $|2\rangle,\,|3\rangle,\cdots$
. These states can naturally be introduced into Eq. (\ref{eq:basic_scattering_S7})
as states pertaining to ``other scattering angles`` as 
\[
|s\rangle\Rightarrow\cos\theta|s\rangle+i\sin\theta|a\rangle+\varepsilon_{2}|2\rangle+\varepsilon_{3}|3\rangle+\cdots,
\]
\begin{equation}
|a\rangle\Rightarrow\cos\theta|a\rangle+i\sin\theta|s\rangle+\eta_{2}|2\rangle+\eta_{3}|3\rangle+\cdots,
\end{equation}
where $\varepsilon_{i}$ and $\eta_{i}$ are \emph{unknown} complex
amplitudes and the right hand side is no longer normalized. Note that,
in the basis $|0\rangle,\,|1\rangle$, this relation is expressed
as 
\[
|0\rangle\Rightarrow e^{i\theta}|0\rangle+\frac{\varepsilon_{2}+\eta_{2}}{\sqrt{2}}|2\rangle+\frac{\varepsilon_{3}+\eta_{3}}{\sqrt{2}}|3\rangle+\cdots,
\]
\begin{equation}
|1\rangle\Rightarrow e^{-i\theta}|1\rangle+\frac{\varepsilon_{2}-\eta_{2}}{\sqrt{2}}|2\rangle+\frac{\varepsilon_{3}-\eta_{3}}{\sqrt{2}}|3\rangle+\cdots,
\end{equation}
which also represents scattering into other states. Now, suppose that
we found the electron in the state $|2\rangle$ due to elastic scattering.
Then, Eq. (\ref{eq:combined_initial2_S7}) is transformed into a Q1
state 
\begin{equation}
\cos\alpha\left(\frac{\varepsilon_{2}|\overline{s}\rangle+\eta_{2}|\overline{a}\rangle}{\sqrt{2}}\right)+i\sin\alpha\left(\frac{\varepsilon_{2}|\overline{a}\rangle+\eta_{2}|\overline{s}\rangle}{\sqrt{2}}\right).
\end{equation}
This is a disaster, from which we cannot recover. One way to avoid
it is to detect the electron in a state that has a component of the
form $|0\rangle+e^{i\theta}|1\rangle$ alongside $|2\rangle,\,|3\rangle,\cdots$
\cite{eeem_cpb,eeem_error}. However, this mixes $|s\rangle$ and
$|a\rangle$, making handling of inelastic scattering difficult. In
the main text, we describe a satisfactory solution to this problem.

\section*{APPENDIX B: DATA ANALYSIS WITH A SHIFTED ELECTRON BEAM ARRAY}

The scheme in the main text uses an array of focused electron beams.
However, a focused electron beams would quickly destroy the specimen
at the focal points. Nonetheless, focused electron beams are suitable
for a proposed quantum electron detector \cite{q_interface} that
could, in principle, transfer the electron state to a quantum information
processor. To solve this problem of specimen damage, we propose to
shift the beam array every time a single round of quantum measurement
is done. Below we describe how to process the data obtained in that
way. Note that, within a single round using $k$ electrons, we need
to focus the beam at the same point, or at least these $k$ focal
points should all be within an area that equals the desired resolution
squared.

We consider a $1$-dimensional case with $M\gg1$, without losing
the gist of the argument. Hence we consider a 1-dimensional map of
phase shift $\theta\left(x\right)$ and we measure 
\begin{equation}
\overline{\theta}=\frac{1}{M}\sum_{-\frac{M}{2}\le n<\frac{M}{2}}\left(-1\right)^{n}\theta_{n},\label{eq:average_theta_S6}
\end{equation}
where $\theta_{n}=\theta\left(n\sigma\right)$.

We show that this measurement detects SFs 
\begin{equation}
q=\frac{\pi}{\sigma},\frac{3\pi}{\sigma},\frac{5\pi}{\sigma},\cdots\label{eq:detectable_k_S6}
\end{equation}
in the case $M\rightarrow\infty$. (Alternatively, the reader may
convince themselves by drawing diagrams.) Define $\delta_{S}\left(x\right)$
as 
\begin{equation}
\delta_{S}\left(x\right)=\sum_{n\in\mathbb{Z}}\delta\left(x-2n\sigma\right)-\sum_{n\in\mathbb{Z}}\delta\left(x-\sigma-2n\sigma\right).
\end{equation}
Then, from Eq. (\ref{eq:average_theta_S6}) we obtain, since $M=\frac{L}{\sigma}$,
\begin{equation}
\overline{\theta}=\lim_{L\rightarrow\infty}\frac{\sigma}{L}\int_{-\frac{L}{2}}^{\frac{L}{2}}\theta\left(x\right)\delta_{S}\left(x\right)dx.
\end{equation}
By Plancherel's theorem, the integral part in the above equals, noting
$\delta_{S}\left(x\right)=\delta_{S}^{*}\left(x\right)$, 
\begin{equation}
\int_{-\infty}^{\infty}\theta\left(x\right)\delta_{S}^{*}\left(x\right)dx=\int_{-\infty}^{\infty}\Theta\left(q\right)\Delta_{S}^{*}\left(q\right)\frac{dq}{2\pi},
\end{equation}
where 
\begin{equation}
\Theta\left(q\right)=\int_{-\infty}^{\infty}\theta\left(x\right)e^{-iqx}dx,
\end{equation}
and 
\[
\Delta_{S}\left(q\right)=\int_{-\infty}^{\infty}\delta_{S}\left(x\right)e^{-iqx}dx
\]
\[
=\left(1-e^{-iq\sigma}\right)\sum_{n\in\mathbb{Z}}\delta\left(\frac{q\sigma}{\pi}-n\right)
\]
\begin{equation}
=\frac{2\pi}{\sigma}\sum_{m\in\mathbb{Z}}\delta\left(q-\frac{\left(2m+1\right)\pi}{\sigma}\right),
\end{equation}
where we used an identity $\sum_{n\in\mathbb{Z}}e^{2\pi inx}=\sum_{n\in\mathbb{Z}}\delta\left(x-n\right)$.
Putting results together, we find 
\begin{equation}
\overline{\theta}\propto\sum_{m\in\mathbb{Z}}\int_{-\infty}^{\infty}\Theta\left(q\right)\delta\left(q-\frac{\left(2m+1\right)\pi}{\sigma}\right)\frac{dq}{2\pi},
\end{equation}
which is what we wanted to show. 

Of the SFs in Eq. (\ref{eq:detectable_k_S6}), virtually only $q=\frac{\pi}{\sigma}$
is important because finer structures are generally smaller in cryoEM
(See Sec. \ref{sec:specimen_damage}). The scheme is insensitive to
all other SFs, as shown above. Hence we focus on the $q=\frac{\pi}{\sigma}$
component, that is of the form: 
\begin{equation}
\theta\left(x\right)=A\cos\left(qx+\phi\right)=A\cos\left(\frac{\pi}{\sigma}x+\phi\right).\label{eq:cos_curve_S6}
\end{equation}
Eq. (\ref{eq:average_theta_S6}) yields, for $\theta_{n}=\theta\left(n\sigma\right)$,
\begin{equation}
\overline{\theta}=A\cos\phi.\label{eq:A_cos_phi_S6}
\end{equation}
To obtain full information, we do another measurement at $\theta_{n}=\theta\left(n\sigma+\frac{\sigma}{2}\right)$
to obtain 
\begin{equation}
\overline{\theta}=-A\sin\phi.\label{eq:A_sin_phi_S6}
\end{equation}
Equations (\ref{eq:A_cos_phi_S6}, \ref{eq:A_sin_phi_S6}) clearly
gives $A$ and $\phi$, which are all the information about the spatial
frequency $q=\frac{\pi}{\sigma}$.

To avoid excessive damaging of the specimen at $x=\frac{n}{2}\sigma$,
where $n\in\mathbb{Z}$, consider measurements at $x=\left(\frac{n}{2}+\frac{\delta}{\pi}\right)\sigma$
for some $\delta$. Equation (\ref{eq:cos_curve_S6}) shows that,
in this case, we can replace $\phi$ with $\phi+\delta$ in all the
calculation above. Thus we obtain 
\begin{equation}
\overline{\theta}=A\cos\left(\phi+\delta\right)\label{eq:A_cos_phi-1_S6}
\end{equation}
for $\theta_{n}=\theta\left(\left(n+\frac{\delta}{\pi}\right)\sigma\right)$
and 
\begin{equation}
\overline{\theta}=-A\sin\left(\phi+\delta\right)\label{eq:A_sin_phi-1_S6}
\end{equation}
for $\theta_{n}=\theta\left(\left(n+\frac{1}{2}+\frac{\delta}{\pi}\right)\sigma\right)$.
Clearly, Eqs. (\ref{eq:A_cos_phi-1_S6}, \ref{eq:A_sin_phi-1_S6})
yield $A$ and $\phi$ as well.

\section*{APPENDIX C: ELECTRON WAVEFUNCTION AFTER INELASTIC SCATTERING}

\subsection{Brief review of inelastic scattering}

Here we review some known facts about inelastic scattering, in part
because we also want to fix notations. See Refs. \cite{Egerton_textbook,Messiah_qm}
for further information. Let $a_{0}$ be Bohr radius $\frac{4\pi\varepsilon_{0}\hbar^{2}}{m_{e}e^{2}}$.
Let $R$ be Rydberg energy $\frac{\hbar^{2}}{2m_{e}a_{0}^{2}}$. Consider
incident electron plane wave $e^{i\mathbf{k}_{i}\cdot\mathbf{r}}=\langle\mathbf{r}|\mathbf{k}_{i}\rangle$
and an outgoing plane wave $e^{i\mathbf{k}_{f}\cdot\mathbf{r}}=\langle\mathbf{r}|\mathbf{k}_{f}\rangle$.
Upon inelastic scattering, the specimen is excited from the ``ground
state'' $|g\rangle$ to an excited state $|e\rangle$. Let the scattering
vector be $\mathbf{q}=\mathbf{k}_{f}-\mathbf{k}_{i}$. Let the Hamiltonian
be $H=H_{p}+H_{0}+V$, where $H_{p}$ is the kinetic energy of the
probe electron, while $H_{0}$ contains kinetic energy of (possibly
multiple) nuclei and electrons in the specimen, \emph{and} the potential
energy describing their interactions. In short, $H_{p}+H_{0}$ is
the non-interacting part in terms of the probe-specimen interaction.
Let the number of relevant electrons involved within the specimen
be $N$. The interaction term $V$, which describes interaction between
the probe electron and the specimen, is 
\begin{equation}
V\left(\mathbf{r}\right)=V_{N}+\frac{e^{2}}{4\pi\varepsilon_{0}}\sum_{i=1}^{N}\frac{1}{\left|\mathbf{r}-\mathbf{r}_{i}\right|},
\end{equation}
where $V_{N}$ describes interaction between the probe electron and
the atomic nuclei; and $\mathbf{r}_{i}$ is the position of $i$-th
electron in the specimen. Let the specimen wavefunction, e.g. the
one pertaining to the ground state $\Psi\left(\mathbf{r}_{1}\mathbf{r}_{2}\cdots\mathbf{r}_{N}\right)$
be anti-symmetrized already. Theory of inelastic scattering tells
us that the differential scattering cross section is 
\begin{equation}
\frac{d\sigma}{d\Omega}=\frac{m_{e}^{2}}{4\pi^{2}\hbar^{4}}\frac{k_{f}}{k_{i}}|\langle f|V|i\rangle|^{2},\label{eq:cross_section}
\end{equation}
where $|i\rangle=|\mathbf{k}_{i}g\rangle$ and $|f\rangle=|\mathbf{k}_{f}e\rangle$.
(Note that dimension of $|\mathbf{k}_{i}\rangle,|\mathbf{k}_{f}\rangle$
is $\mathrm{L}^{\frac{3}{2}}$ because they are normalized as $\langle\mathbf{k}_{f}|\mathbf{k}_{i}\rangle=\left(2\pi\right)^{3}\delta^{2}\left(\mathbf{k}_{f}-\mathbf{k}_{i}\right)$.)
The same quantity is often expressed using the \emph{generalized oscillator
strength} (GOS) $f\left(q\right)$ as 
\begin{equation}
\frac{d\sigma}{d\Omega}=\frac{4\gamma^{2}R}{Eq^{2}}\frac{k_{f}}{k_{i}}f\left(q\right).\label{eq:cross_section_GOS}
\end{equation}
The GOS is known to reduce to the \emph{dipole oscillator strength}
when $q\rightarrow0$. (The question of ``Compared to what?'' will
be answered shortly.) Equation (\ref{eq:cross_section_GOS}) is used
not only for excitations of inner shell electrons or that of isolated
atoms, but also in the case of outer-shell excitations, where chemical
bondings between atoms play a role, and collective excitations such
as plasmons. In all these cases, $f\left(q\right)$ tends to have
a constant value in the \emph{dipole region} \cite{Egerton_textbook}.
Henceforth we assume that most of relevant scattering is in the dipole
region. In this case, GOS is known to be expressed as 
\begin{equation}
f\left(q\right)=\frac{E}{R}\frac{\left|\varepsilon\left(\mathbf{q}\right)\right|^{2}}{\left(qa_{0}\right)^{2}},\label{eq:GOS}
\end{equation}
where $\varepsilon\left(\mathbf{q}\right)$ is the \emph{inelastic
form factor}. The dimensionless form factor $\varepsilon\left(\mathbf{q}\right)$
is given as 
\begin{equation}
\varepsilon\left(\mathbf{q}\right)=\sum_{i=1}^{N}\langle e|e^{i\mathbf{q}\cdot\mathbf{r}_{i}}|g\rangle=N\langle e|e^{i\mathbf{q}\cdot\mathbf{r}_{1}}|g\rangle.
\end{equation}
The second equality holds because all electrons are identical particles
and equivalent. Expanding this, we obtain 
\begin{equation}
N\langle e|e^{i\mathbf{q}\cdot\mathbf{r}_{1}}|g\rangle=N\langle e|g\rangle+Ni\mathbf{q}\cdot\langle e|\mathbf{r}_{1}|g\rangle+\cdots.
\end{equation}
The first, zeroth-order term vanishes by orthogonality. We assume
that the spatial extent of relevant bound electron states are small
compared to $\frac{2\pi}{q}$, so that the second and higher-order
terms are negligible, which is another way to say that we work in
the dipole region. Hence we obtain 
\begin{equation}
\varepsilon\left(\mathbf{q}\right)=\mathbf{q}\cdot\mathbf{a},
\end{equation}
where $\mathbf{a}=Ni\langle e|\mathbf{r}_{1}|g\rangle$.

\subsection{Assumption about dipole-region scattering}

The vector $\mathbf{a}=Ni\langle e|\mathbf{r}_{1}|g\rangle$ may have
real and imaginary parts, as in 
\begin{equation}
\mathbf{a}=\mathbf{a}_{R}+i\mathbf{a}_{I},
\end{equation}
where both $\mathbf{a}_{R}$ and $\mathbf{a}_{I}$ are real vectors
with unknown directions and lengths. At this point, we make a second
assumption that $\mathbf{a}_{R}$ and $\mathbf{a}_{I}$ are parallel
to each other. Then, we can regard the vector $\mathbf{a}$ simply
as a real vector, up to an unimportant overall phase factor $e^{iu}$
that we omit hereafter. To visualize the meaning of this assumption,
suppose, instead, that $\mathbf{a}_{R}\propto\hat{\mathbf{i}}$ and
$\mathbf{a}_{I}\propto\hat{\mathbf{j}}$. We will later see that the
scattered electron wave in the far field is essentially $\frac{\varepsilon\left(\mathbf{q}\right)}{q^{2}}$.
We also note that relevant $\mathbf{q}$ approximately lies within
the $xy$ plane. We find that the above wavefunction 
\begin{equation}
\frac{\varepsilon\left(\mathbf{q}\right)}{q^{2}}=\frac{\mathbf{q}\cdot\mathbf{a}_{R}+i\mathbf{q}\cdot\mathbf{a}_{I}}{q^{2}}
\end{equation}
cannot be superposed onto its own mirror image, unless we ``peel
the wavefunction off the $xy$ plane''. The word ``dipole region''
feels inappropriate for this kind of state, which we may call \emph{chiral},
although there appears to be no such definitions in the literature,
presumably because only the statistical average of the square of wavefunctions
mattered thus far. Hence our second assumption is that the scattered
electron state is \emph{achiral }in the above sense.

\subsection{The exit wave after inelastic scattering\label{subsec:The-exit-wave_S8}}

Consider the wavefunction of the probe electron right after inelastic
scattering. Let the time evolution operator be $U\left(t\right)$.
Noting that $\langle\mathbf{k}_{f}e|\Psi_{\mathrm{final}}\rangle$
is the wavefunction of the final state in the reciprocal space $\Psi_{0}\left(\mathbf{k}_{f}\right)$,
we intend to find, for a large $t$, 
\begin{equation}
\Psi_{0}\left(\mathbf{k}_{f}\right)=\langle\mathbf{k}_{f}e|\Psi_{\mathrm{final}}\rangle=\langle\mathbf{k}_{f}e|U\left(t\right)|\mathbf{k}_{i}g\rangle=\langle f|U\left(t\right)|i\rangle.
\end{equation}
On the other hand, the following expression appears in standard derivations
of Fermi's golden rule: 
\[
\langle f|U\left(t\right)|i\rangle=-\frac{\langle f|V|i\rangle}{\hbar\omega}\left(e^{i\omega t}-1\right)
\]
\begin{equation}
=-\frac{it}{\hbar}\langle f|V|i\rangle e^{i\frac{\omega t}{2}}\frac{\sin\left(\frac{\omega t}{2}\right)}{\frac{\omega t}{2}}.
\end{equation}
It is also known that (The reader may convince themselves, using contour
integration etc.) 
\begin{equation}
k\frac{\sin kx}{kx}\xrightarrow{{k\rightarrow\infty}}\pi\delta\left(x\right).
\end{equation}
Using this, the above expression is modified to 
\begin{equation}
\langle f|U\left(t\right)|i\rangle\xrightarrow{{t\rightarrow\infty}}-\frac{2\pi i}{\hbar}\langle f|V|i\rangle e^{i\frac{\omega t}{2}}\delta\left(\omega\right).
\end{equation}
\emph{If} we restrict the range of $\mathbf{k}_{f}$ to ones that
satisfy energy conservation of the inelastic scattering process, we
can omit the factor $\delta\left(\omega\right)$ to obtain, neglecting
the unimportant proportional factor 
\begin{equation}
\langle f|U\left(t\right)|i\rangle\xrightarrow{{t\rightarrow\infty}}\langle f|V|i\rangle.\label{eq:t_inf_limit_S8}
\end{equation}
Hence we obtain 
\[
\Psi_{0}\left(\mathbf{k}_{f}\right)\propto\langle f|V|i\rangle=\langle\mathbf{k}_{f}e|V|\mathbf{k}_{i}g\rangle.
\]
We find, using Eqs. (\ref{eq:cross_section},\ref{eq:cross_section_GOS},\ref{eq:GOS}),
this is proportional to 
\begin{equation}
\frac{\varepsilon\left(\mathbf{q}\right)}{q^{2}}\propto\frac{\mathbf{q}\cdot\mathbf{a}}{q^{2}}.
\end{equation}
Recall that $\mathbf{q}$ is a shorthand for $\mathbf{k}_{f}-\mathbf{k}_{i}$.
We write the energy of the incident electron as 
\begin{equation}
E_{K}=\sqrt{m_{e}^{2}c^{4}+c^{2}p^{2}}-m_{e}c^{2}=E_{R}-m_{e}c^{2},
\end{equation}
where we also defined $E_{R}$. In terms of the scattering angle $\theta$,
measured from the original direction $\mathbf{k}_{i}$, standard considerations
\cite{Egerton_textbook} yield the wavefunction 
\begin{equation}
\psi\left(\mathbf{k}_{f}\right)\propto\frac{1}{q}\left(\frac{\mathbf{q}}{q}\right)\cdot\mathbf{a}=\frac{1}{k_{i}\sqrt{\theta^{2}+\theta_{E}^{2}}}\left(\frac{\mathbf{q}}{q}\right)\cdot\mathbf{a},
\end{equation}
where $\theta_{E}=\frac{\Delta k}{k_{i}}\approx\frac{E}{2E_{K}}$.
Its numerical value is $\theta_{E}=41\,\mu\mathrm{rad}$ for $E_{K}=300\,\mathrm{keV}$
and $E=20\,\mathrm{eV}$. This is very small compared to typical scattering
angles and hence variation of $E$ affects only the region $\theta\lesssim\theta_{E}$.
This is why we mentioned, at Sec. \ref{subsec:Preliminary-remarks},
that the scattered electron state only weakly depends on the final
state $|e\rangle$ of the specimen, thus justifying our not using
mixed quantum states.

At a larger scattering angle, we impose a cut off to $\Psi_{0}\left(\mathbf{k}_{f}\right)$
at the Bethe ridge at the scattering angle $\theta_{c}$. Suppose
that an incident electron, with energy $E_{K}$, knocks a single bound
electron off its bound state with a binding energy $E_{B}$. Purely
kinematic considerations on energy and momentum conservation, where
we assume that the bound electron remains to be non-relativistic after
being knocked off the bound state, tells us that the probe electron
undergoes scattering with a scattering angle $\theta$, which reaches
the maximum $\theta_{c}=\sqrt{\frac{2\theta_{E}}{\gamma}}$ with respect
to $E_{B}$, when $E_{B}=0$.

We make a brief digression to derive the relation $\theta_{c}=\sqrt{\frac{2\theta_{E}}{\gamma}}$.
From the momentum conservation 
\[
q^{2}=k_{i}^{2}+k_{f}^{2}-2k_{i}k_{f}\cos\theta
\]
\begin{equation}
\approx\left(k_{i}-k_{f}\right)^{2}+k_{i}k_{f}\theta^{2}\approx\Delta k^{2}+k^{2}\theta^{2},\label{eq:law_cosine}
\end{equation}
where $\Delta k=\left|k_{i}-k_{f}\right|$. We also have energy conservation
\begin{equation}
\sqrt{m_{e}^{2}c^{4}+c^{2}\hbar^{2}k_{i}^{2}}=\sqrt{m_{e}^{2}c^{4}+c^{2}\hbar^{2}k_{f}^{2}}+\frac{\hbar^{2}q^{2}}{2m_{e}}+E_{B}.
\end{equation}
Note that the energy loss $E$ is 
\[
E=\sqrt{m_{e}^{2}c^{4}+c^{2}\hbar^{2}k_{i}^{2}}-\sqrt{m_{e}^{2}c^{4}+c^{2}\hbar^{2}k_{f}^{2}}
\]
\begin{equation}
=\frac{dE}{dp}\hbar\Delta k=\frac{\hbar^{2}c^{2}k}{E_{R}}\Delta k.
\end{equation}
Hence, energy conservation is simplified to $\frac{k}{E_{R}}\Delta k=\frac{q^{2}}{2m_{e}c^{2}}+\frac{E_{B}}{\hbar^{2}c^{2}}$,
or 
\begin{equation}
2k\Delta k=\gamma q^{2}+\frac{2E_{R}E_{B}}{\hbar^{2}c^{2}}.
\end{equation}
Combining this with the momentum conservation relation, we obtain
\begin{equation}
\frac{2k\Delta k}{\gamma}-\frac{2E_{R}E_{B}}{\gamma\hbar^{2}c^{2}}=q^{2}=\Delta k^{2}+k^{2}\theta^{2},
\end{equation}
and hence 
\begin{equation}
\theta^{2}=\frac{2\Delta k}{\gamma k}-\left(\frac{\Delta k}{k}\right)^{2}-\frac{2E_{R}E_{B}}{\gamma k^{2}\hbar^{2}c^{2}}.
\end{equation}
The angle $\theta$ reaches the maximum $\theta_{c}$ with respect
to the binding energy $E_{B}$, when $E_{B}=0$: 
\begin{equation}
\theta^{2}<\theta_{c}^{2}=\frac{2\Delta k}{\gamma k}-\left(\frac{\Delta k}{k}\right)^{2}\approx\frac{2\Delta k}{\gamma k}.
\end{equation}
Recalling $\frac{\Delta k}{k}=\theta_{E}$, we obtain the Bethe ridge
angle 
\begin{equation}
\theta_{c}=\sqrt{\frac{2\theta_{E}}{\gamma}},\label{eq:bethe_ridge}
\end{equation}
which is $7.2\,\mathrm{mrad}$ for $300\,\mathrm{keV}$ electrons.

Summarizing, the electron that underwent inelastic scattering has
a wavefunction in the far field: 
\begin{equation}
\Psi_{0}\left(\mathbf{q}\right)=\begin{cases}
\frac{1}{\sqrt{\theta^{2}+\theta_{E}^{2}}}\left(\frac{\mathbf{q}}{q}\right)\cdot\mathbf{a} & \theta<\theta_{c}\\
0 & \theta>\theta_{c}
\end{cases}\label{eq:scattered_wavefunction}
\end{equation}
where $\mathbf{a}$ has an unknown direction, and we write the wavefunction
as a function of $\mathbf{q}=\mathbf{k}_{f}-\mathbf{k}_{i}$ rather
than that of the final momentum $\mathbf{k}_{f}$ for later convenience.
The magnitude of $\mathbf{a}$ is unimportant because it is absorbed
in the overall normalization factor. Since in most cases $\mathbf{q}$
is approximately in the $xy$ plane, only $x,y$ components of $\mathbf{a}$
is important. 

Although Eq. (\ref{eq:t_inf_limit_S8}) is $t\rightarrow\infty$ limit,
since the electron propagates in the free space after scattering,
we should be able to find the wave function $\psi_{0}\left(\mathbf{r}\right)$
right after scattering by simply performing inverse Fourier transform
to $\Psi_{0}\left(\mathbf{q}\right)$: 
\begin{equation}
\psi_{0}\left(\mathbf{r}\right)=\mathcal{F}_{C}^{-1}\left\{ \Psi_{0}\left(\mathbf{q}\right)\right\} .
\end{equation}

Equation (\ref{eq:scattered_wavefunction}) assumed that the scattering
occurred at the origin $x=y=0$. Since actual inelastic scattering
occurs at an unknown location $\mathbf{r}_{0}$, we need to generalize
this result. Fourier transforming $\psi_{0}\left(\mathbf{r}-\mathbf{r}_{0}\right)$
suffices for this purpose. Thus we obtain 
\[
\Psi_{1}\left(\mathbf{q}\right)=\int\psi_{0}\left(\mathbf{r}-\mathbf{r}_{0}\right)e^{-i\mathbf{q}\cdot\mathbf{r}}d^{2}\mathbf{r}
\]

\begin{equation}
=e^{-i\mathbf{q}\cdot\mathbf{r}_{0}}\int\psi_{0}\left(\mathbf{r}-\mathbf{r}_{0}\right)e^{-i\mathbf{q}\cdot\left(\mathbf{r}-\mathbf{r}_{0}\right)}d^{2}\mathbf{r}=e^{-i\mathbf{q}\cdot\mathbf{r}_{0}}\Psi_{0}\left(\mathbf{q}\right)
\end{equation}

\section*{APPENDIX D: COMPUTING THE PHASE MAP}

We computed the phase map (Fig. 4(a)) of the Marburg virus VP35 oligomerization
domain (5TOI) using the multislice algorithm. The thickness of each
slice is $1\,\mathrm{nm}$. A simpler simulation using the projection
assumption \cite{Rez} gave very similar results, which is not surprising
because the thickness of the 5TOI molecule is as thin as $\approx3\,\mathrm{nm}$.

\begin{table}[t]
\caption{The inner potential $V_{i}$ (multiplied by the volume).}
\begin{tabular}{|c||c|c|c|c|c|}
\hline 
Element  & H  & C  & N  & O  & S\tabularnewline
\hline 
\hline 
$V_{i}/\mathrm{V\,nm^{3}}$  & 0.0253  & 0.118  & 0.106  & 0.095  & 0.246\tabularnewline
\hline 
\end{tabular}
\end{table}

\begin{table}[t]
\caption{Atomic radii $a_{i}$.}
\begin{tabular}{|c||c|c|c|c|c|}
\hline 
Element  & H  & C  & N  & O  & S\tabularnewline
\hline 
\hline 
$a_{i}/\mathrm{nm}$  & 0  & 0.180  & 0.164  & 0.144  & 0.177\tabularnewline
\hline 
\end{tabular}
\end{table}

The handling of water molecules surrounding the 5TOI molecule closely
followed the method described by Shang and Sigworth \cite{Shang_Sigworth}.
Here we only describe places where we made deviations from their method
when we took the surrounding water molecules into account. Following
the main text, we focus on $300\,\mathrm{keV}$ electrons. All computations
were carried out on a Cartesian grid with a grid spacing $0.05\,\mathrm{nm}$.
The shape of the space was cubic with the volume $V=L^{3}$, where
$L=12.0\,\mathrm{nm}$.

First, we remark that the surrounding water structure is not \emph{obviously}
averaged out under the assumption of single image acquisition in the
present work, unlike in the context of SPA considered in Ref. \cite{Shang_Sigworth}.
However, there is evidence that water molecules move significantly
during the electron exposure \cite{water_motion}. Here we assume
that the use of averaged-out water density is justified.

We computed the inner potentials of relevant elements H, C, N, O and
S as follows. The scattering amplitudes $f\left(\theta\right)$ at
$\theta=0$ for the elements were obtained from a NIST database \cite{NIST_database}.
From these values we computed the values of inner potentials $V_{i}$
(which has the dimension of voltage times volume) as 
\begin{equation}
V_{i}=\frac{2\pi\hbar^{2}}{\gamma m_{e}e}f\left(0\right).
\end{equation}
Table I shows the result.

The mean inner potential of ice is computed to be $4.5276\,\mathrm{V}$.
In other words, this value represents the inner potential of the water
molecule, consisting of $2$ hydrogen atoms and one oxygen atom, divided
by its molecular volume in ice. There is a discrepancy in the literature
regarding the exact value of it. Reference \cite{Shang_Sigworth}
reports $3.6\,\mathrm{V}$ for ``bulk vitreous ice'', whereas Ref.
\cite{vulovic_etal} reports a value $4.5301\,\mathrm{V}$ for ``low-density
amorphous ice'' (LDA ice). Under the assumption that the density
of LDA ice $9.3\times10^{2}\,\mathrm{kg/m^{3}}$ is relevant, the
latter value, which is consistent with our result, is more appropriate.

The ``atomic radii'' used for computing the ``binary mask function''
$m\left(\mathbf{r}\right)$ \cite{Shang_Sigworth} are shown in Table
II. We use van der Waals (VDW) radii taken from Table 2 of Ref. \cite{Van_der_Waals_radii}
for this purpose. To be precise, the VDW radii depend on the atomic
group to which the atom belongs. However, we simply averaged all values
appearing in the ``ProtOr Radii'' column of the Table 2 of Ref.
\cite{Van_der_Waals_radii}. This is clearly a crude approximation
but we believe that the associated error is insignificant for the
present purpose of evaluating QEM.

Hydrogen requires a special treatment. Atomic coordinates for the
hydrogen atoms are absent in the PDB data, for the Marburg virus VP35
oligomerization domain (5TOI) \cite{5TOI}. Following the general
strategy described in Ref. \cite{Shang_Sigworth}, we modified the
inner potential values of C, N, O and S atoms in accordance with the
expected number of the associated H atoms to each of these elements.
We computed the expected values as weighted-average of the number
of hydrogen atoms in each type of amino-acid residue, over all residue
types with weights in accordance with the frequency of each residue
in the 5TOI molecule.

\end{document}